\documentclass[aps,prb,showpacs,preprintnumbers,amsmath,amssymb,superscriptaddress,twocolumn]{revtex4-1}
\usepackage{graphicx}
\usepackage{comment}
\usepackage{hyperref}
\usepackage{braket}
\usepackage{CJK}

\newcommand{\bs}{\boldsymbol}

\usepackage{xcolor}

\graphicspath{{./figures/}}

\newcommand{\up}{\uparrow}
\newcommand{\down}{\downarrow}

\usepackage{esint}
\makeatletter
\@for\next:={int,iint,iiint,iiiint,dotsint,oint,oiint,sqint,sqiint,
  ointctrclockwise,ointclockwise,varointclockwise,varointctrclockwise,
  fint,varoiint,landupint,landdownint}\do{    \expandafter\edef\csname\next\endcsname{      \noexpand\DOTSI
      \expandafter\noexpand\csname\next op\endcsname
      \noexpand\ilimits@
    }  }
\makeatother

\begin{document}
\title{Quantum Monte Carlo Compton profiles of solid and liquid lithium}
\author{Yubo Yang}
\affiliation{Department of Physics, University of Illinois, Urbana, Illinois 61801, USA}
\author{Nozomu Hiraoka}
\affiliation{National Synchrotron Radiation Research Center, Hsinchu 30076, Taiwan}
\author{Kazuhiro Matsuda}
\affiliation{Graduate School of Science, Kyoto University, Kyoto 606-8502, Japan}
\author{Markus Holzmann}
\affiliation{Univ. Grenoble Alpes, CNRS, LPMMC, 38000 Grenoble, France}
\affiliation{Institut Laue Langevin, BP 156, F-38042 Grenoble Cedex 9, France}
\author{David M. Ceperley}
\affiliation{Department of Physics, University of Illinois, Urbana, Illinois 61801, USA}
\date{\today}
\begin{abstract}
We computed the Compton profile of solid and liquid lithium using quantum Monte Carlo (QMC) and compared with recent experimental measurements obtaining good agreement. Importantly, we find it crucial to account for proper core-valence orthogonalization and to address density differences when comparing with experiment. To account for disorder effects, we sampled finite-temperature configurations using molecular dynamics (MD), then performed diffusion Monte Carlo (DMC) simulations on each configuration. We used Slater-Jastrow wavefunctions and grand-canonical twist-averaged boundary conditions. A QMC pseudopotential correction, derived from an all-electron DMC simulation of the perfect crystal was also used. Our calculations provide the first all-electron QMC benchmark for the Compton profile of lithium crystal and pseudopotential-corrected QMC Compton profiles for both the liquid and solid.
\end{abstract}
\maketitle

\section{Introduction} \label{sec:intro}

The Compton profile is a bulk-sensitive probe of the electronic structure of a material accessible to both theory and experiment. Using the ``impulse approximation''~\cite{Eisenberger1970}, the double differential cross section of inelastic light scattering is directly proportional to the Compton profile, the Radon transform of the electronic momentum distribution along the scattering vector. \begin{equation}
J(p_z) = \iint dk_x dk_y ~ n(k_x, k_y, k_z=p_z),
\end{equation}
where $n(\bs{k})$ is the electronic momentum distribution.
Since the pioneering work of Eisenberger et al.~\cite{Eisenberger1970,Eisenberger1972}, Compton scattering experiments have been performed on simple metals such as Li~\cite{Sakurai1995, Schulke1996,Chen1999,Sternemann2001,Tanaka2001}, Be~\cite{Hamalainen1996,Huotari2000}, Na~\cite{Huotari2010} as well as more complicated materials. Accompanying the scattering experiments are numerous theoretical calculations using different electronic structure theories including density functional theory (DFT)~\cite{Sakurai1995,Schulke1996,Tanaka2001,Jarlborg1998,Baruah1999,Bross2005,Makkonen2005,Klevak2016,Sekania2018}, QMC~\cite{Filippi1999,Huotari2010}, and GW~\cite{Yasunori1997,Schulke1999,Eguiluz2000,Olevano2012}.
The Compton profiles in ref.~\cite{Sakurai1995,Schulke1996} were compared to DFT results using the local density approximation (LDA) with the Lam-Platzman correlation correction~\cite{Lam1974}.
While the Lam-Platzman correction has been shown to be accurate by QMC~\cite{Filippi1999,Schulke2001,Bross2005}, the theoretical Compton profile is still larger at low momenta and smaller at high momenta compared with experiment. In other words, the predicted Compton profile is typically narrower than observed.

Both theoretical approximations and experimental procedures may be responsible for a significant fraction of the aforementioned discrepancy. In the experiment, finite momentum resolution and final-state effects~\cite{Sternemann2000,Soininen2001} broaden the measured Compton profile. In the theoretical calculations, the lack of electronic correlation and the use of pseudopotentials both narrow the computed Compton profile. Furthermore, many subtle complications may also be responsible for part of the discrepancy. Examples include: multiple scattering corrections, background subtraction, thermal expansion, electron-phonon coupling, and relativistic effects.

In this paper, we present much improved QMC calculations on the solid and liquid states of lithium. Firstly, we use grand-canonical twist-averaging~\cite{Lin2001,Holzmann2016} to access the momentum distribution at arbitrary momentum while preserving a sharp Fermi surface. We obtain a momentum resolution of 0.040 a.u., which is higher than the 0.068 a.u. achieved previously~\cite{Filippi1999} (It is straight-forward to further increase momentum resolution given more computational resources). Secondly, we perform diffusion Monte Carlo (DMC) to remove effects of the trial wavefunction. Thirdly, we use all-electron QMC to explore the pseudopotential bias in the Compton profile. We find that the pseudopotential bias is responsible for the majority of discrepancy between pseudopotential QMC and experimental Compton profiles away from the Fermi surface. Fourth and finally, we apply finite-size corrections~\cite{Holzmann2009,Holzmann2011} to obtain the momentum distribution in the thermodynamic limit. Using these improved procedures, we calculate the disorder-averaged Compton profiles for polycrystal and liquid lithium and obtain good agreement with recent high-resolution synchrotron experiment~\cite{Nozomu2019}.

This paper is organized as follows. In section \ref{sec:method}, we describe the simulation methods used to obtain the QMC momentum distributions. In section \ref{sec:results}, we show the QMC momentum distributions and the resulting Compton profiles in comparison with experiment. In section \ref{sec:discussion}, we discuss the influence of various physical effects on the momentum distribution in an attempt to explain the remaining discrepancy between QMC and experiment.

\section{Method} \label{sec:method}

Full-core and pseudopotential QMC calculations have been performed on both  the perfect crystal and disordered lithium configurations. We use Slater-Jastrow trial wavefunction
\begin{align}
\Psi_T = D^{\up} D^{\down} \exp\left[ -\sum\limits_{i<j}^{N} u(\bs{r}_i-\bs{r}_j) - \sum\limits_{i=1}^N \chi(\bs{r}_i) \right],\label{eq:sj}
\end{align}
where $u(\bs{r})$ is the electron-electron Jastrow pair function, $\chi(\bs{r})$ is the electron-ion Jastrow pair function and $\bs{r}_i$ is the position of the i$^{th}$ electron. The Slater determinant $D^{\up/\down}$ is composed of single-particle orbitals obtained using Kohn-Sham (KS) DFT with the LDA functional. In the full-core calculation, we remove the approximate electron-ion cusp from the orbitals and re-introduce the exact cusp condition in the Jastrow function~\cite{Ceperley1981}. The electron-ion Jastrow pair function is split into a sum of core and valence pieces. A flexible Bspline with 16 adjustable knots is used for the core piece ($r<2$ bohr). An electron-electron-ion three-body Jastrow is also added to further improve the all-electron wavefunction. In the pseudopotential calculation, we treat the lithium atoms as pseudo ions of charge +1. The core, screened by 1s electrons, is replaced by the BFD pseudopotential~\cite{Burkatzki2007}.  The electron-electron Jastrow pair function is expressed as a sum of real-space and reciprocal-space parts to accurately describe long-range plasmon fluctuations.

In variational Monte Carlo (VMC), we sample $\vert \psi_T \vert^2$ using Metropolis Monte Carlo and directly calculate properties from the many-body wavefunction. The momentum distribution is calculated using the direct estimator in reciprocal space\cite{McMillan1965}. In DMC, an ensemble of electron configurations evolve according to the Green's function of the non-relativistic Schr\"odinger equation in imaginary time. Using the trial wavefunction $\psi_T$ as guiding function and phase reference, the long-time solution samples the mixed distribution $\psi^*_T\psi_{FP}$, in the limit of small time step. $\psi_{FP}$ is the fixed-phase ground-state wavefunction. If the phase of $\psi_T$ were exact, then $\psi_{FP}$ would be the exact ground-state wavefunction.~\cite{Ortiz1993} The difference between the expectation value of an observable in the fixed-phase and the mixed distributions is the mixed-estimator bias. We gauge simulation quality by monitoring kinetic, potential, and total energies as well as pair correlation functions and the momentum distribution. We observe fast equilibration, small variance and small mixed-estimator bias in all monitored quantities. The DMC momentum distribution is linearly extrapolated to remove the mixed-estimator bias. For more details on the computational methods and data processing, see the supplementary materials.

We use GCTABC to improve the momentum distribution \cite{PhysRevLett.97.076404,Holzmann2009}. A previous QMC calculation~\cite{Filippi1999} used real wavefunctions and canonical twist average boundary condition (CTABC); each boundary condition (twist) had the same number of electrons. Use of real trial functions restricted the accessible momenta to those commensurate with the simulation cell. CTABC can occupy states outside of the Fermi surface at certain twists, which artificially smears the Fermi surface. In contrast, the grand-canonical twist average technique enforces constant chemical potential at all twists. We adjust the number of electrons at each twist such that no state outside the Fermi surface is occupied. This allows us to sample the momentum distribution at momenta arbitarily close to the Fermi surface while maintaining a sharp Fermi surface. In practice, we impose the occupation of the orbitals in the Slater determinant according to the LDA Fermi energy. In principle, one might modify the Fermi surface by estimating the chemical potential directly within QMC \cite{yang2019electronic}. However, this is much more computationally demanding and is beyond the scope of the current study and not thought to be necessary for lithium.

In the perfect crystal, the full-core simulation contains 54 lithium atoms, while the pseudopotential simulations contain 54 or 432 atoms. We use MD with the modified embedded-atom potential (MEAM)~\cite{Baskes1992} to generate the disordered configurations.
The MD temperatures were elevated to model quantum fluctuations of the nuclei~\cite{Filippi1998}.
We sample the canonical distribution with 432 lithium atoms at 330K and 500K for experiments at 298K and 493K, respectively.

All calculations have been performed at the same density $r_s=3.25$, consistent with the previous QMC study~\cite{Filippi1999}. After obtaining QMC results at $r_s=3.25$, we rescale the density of QMC Compton profiles to match the experimental densities: $r_s=3.31$ for the liquid and $r_s=3.265$ for the solid.

In both QMC and experiment, we assume the momentum distribution of the core electrons to remain unmodified from that in the isolated atom. The atomic core orbital is calculated using Hartree-Fock (HF) and removed from all-electron results to produce valence electron contributions.

We convolved our QMC Compton profile with a broadening function to model instrument resolution and final-state interaction. For this we used the \emph{extended Lorentzian}
\begin{equation}
b(x) = \frac{1}{\tilde{\Omega}} \frac{1}{
a_0+a_1(\frac{2x}{\Gamma})^2+a_2(\frac{2x}{\Gamma})^4
}\label{eq:elorentz}
\end{equation}
with $\Gamma=0.024$ a.u., $a_0=1$, $a_1=0.85$ and $a_2=0.15$ chosen to fit the convolution of the elastic line in the X-ray experiment and the spectral density function of the electrons and $\tilde{\Omega}$ such that $\int dx ~b(x)=1$. 

We used LAMMPS~\cite{Plimpton1995} for the MD simulations, QE~\cite{Giannozzi2009,Giannozzi2017} for DFT, PySCF~\cite{PYSCF} for HF, and QMCPACK~\cite{Kim2018} for QMC. The disordered calculations have been automated using the nexus suite of tools~\cite{Krogel2016}.

\section{Results} \label{sec:results}

\begin{figure}[h]
\includegraphics[width=\linewidth]{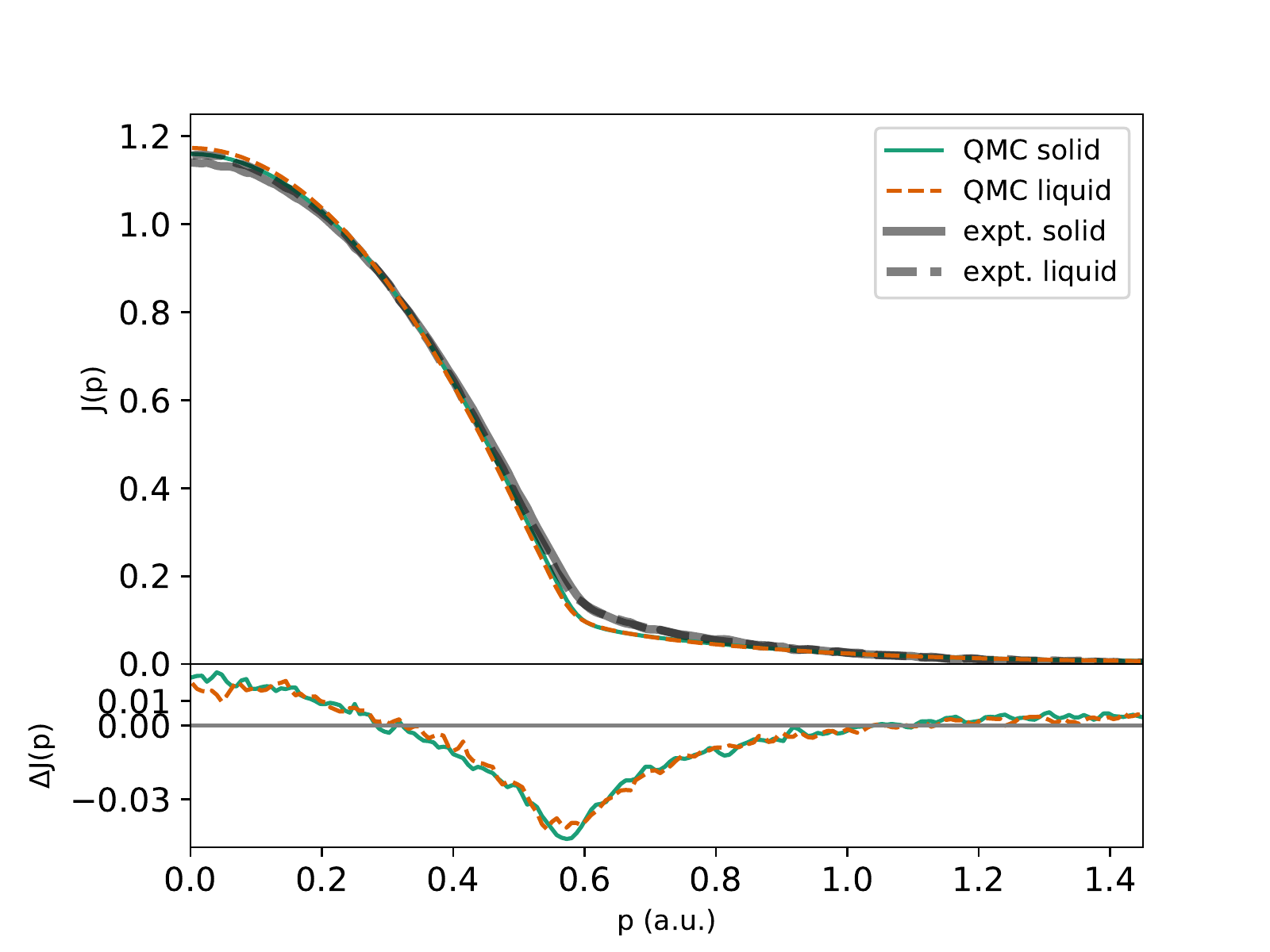}
\caption{Valence electronic Compton profiles of solid (solid line) and liquid (dashed line) lithium from QMC (thin) and experiment (thick). The top panel shows the Compton profiles on an absolute scale. The bottom panel shows  $\Delta J(p)=J_{QMC}-J_{expt}$.}\label{fig:sl-jp-djp}
\end{figure}

Figure~\ref{fig:sl-jp-djp} shows the valence Compton profiles of solid and liquid lithium from experiment and processed QMC data. The raw QMC data have been processed to account for finite-size effects, thermal disorder, pseudopotential bias, density change, final-state effects, and instrument resolution. The QMC Compton profiles agree with experiment immediately inside the Fermi surface (0.2 a.u.$<$p$<$0.4 a.u.) and at large momenta (p$>$0.9  a.u.). However, the QMC Compton profiles show less high-momentum component immediately outside the Fermi surface and too much low-momentum component. Both the theoretical and experimental valence Compton profiles satisfy the normalization sum rule ($\int_{-\infty}^{\infty} J(p)dp=1$) to better than 0.3\%. The difference between QMC and experiment Compton profiles can be interpreted as a shift of momentum density from zero to slightly above the Fermi momentum. 

\begin{figure}[h]
\includegraphics[width=\linewidth]{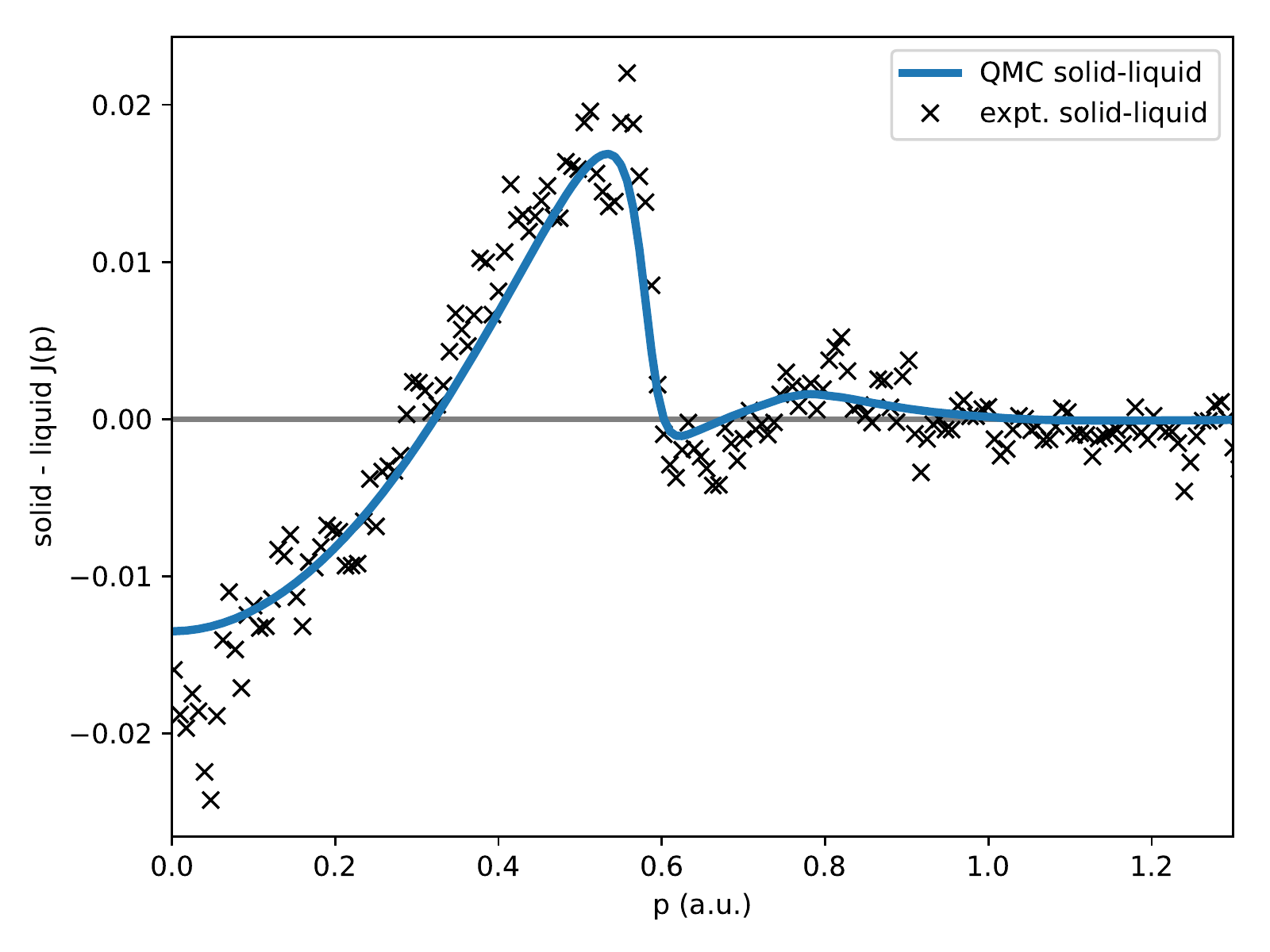}
\caption{Difference between solid and liquid valence electronic Compton profiles.\label{fig:s-l-djp}}
\end{figure}

Figure~\ref{fig:s-l-djp} shows the change of the Compton profile when the liquid freezes into a solid. The systematic difference  between QMC calculations and experiment is almost identical in the solid and liquid. Thus, cancellation of error allows us to capture the difference between the solid and liquid Compton profiles almost perfectly.
The main change is a density-induced outward shift of the Fermi surface. This shift manifests in Fig.~\ref{fig:s-l-djp} as a peak at the solid Fermi momentum $p_F\approx0.578$ a.u. and a parabolic dip centered around $p=0$. Another important difference is the emergence of secondary Fermi surfaces, due to Umklapp scattering in the solid. We expect secondary Fermi surfaces to center around the reciprocal lattice of the lithium crystal. Crystalline lithium is BCC with a lattice constant of $\sim 6.63$ bohr, so its reciprocal lattice is FCC with a lattice constant of $\sim 1.895$ a.u.. The nearest neighbor to $\Gamma$ is $p_1=1.34$ a.u.  along [110]. Therefore, the closest secondary Fermi surface is located at $p_1-p_F=0.762$ a.u., which is exactly where we observe a small peak in Fig.~\ref{fig:s-l-djp}.

As mentioned at the beginning of this section, we process the raw QMC data in several steps to make them comparable to experiment. In the following, we present perfect lithium crystal QMC calculations, which we use to validate the processing steps.

\begin{figure}[h]
\begin{minipage}{\columnwidth}
\includegraphics[width=\linewidth]{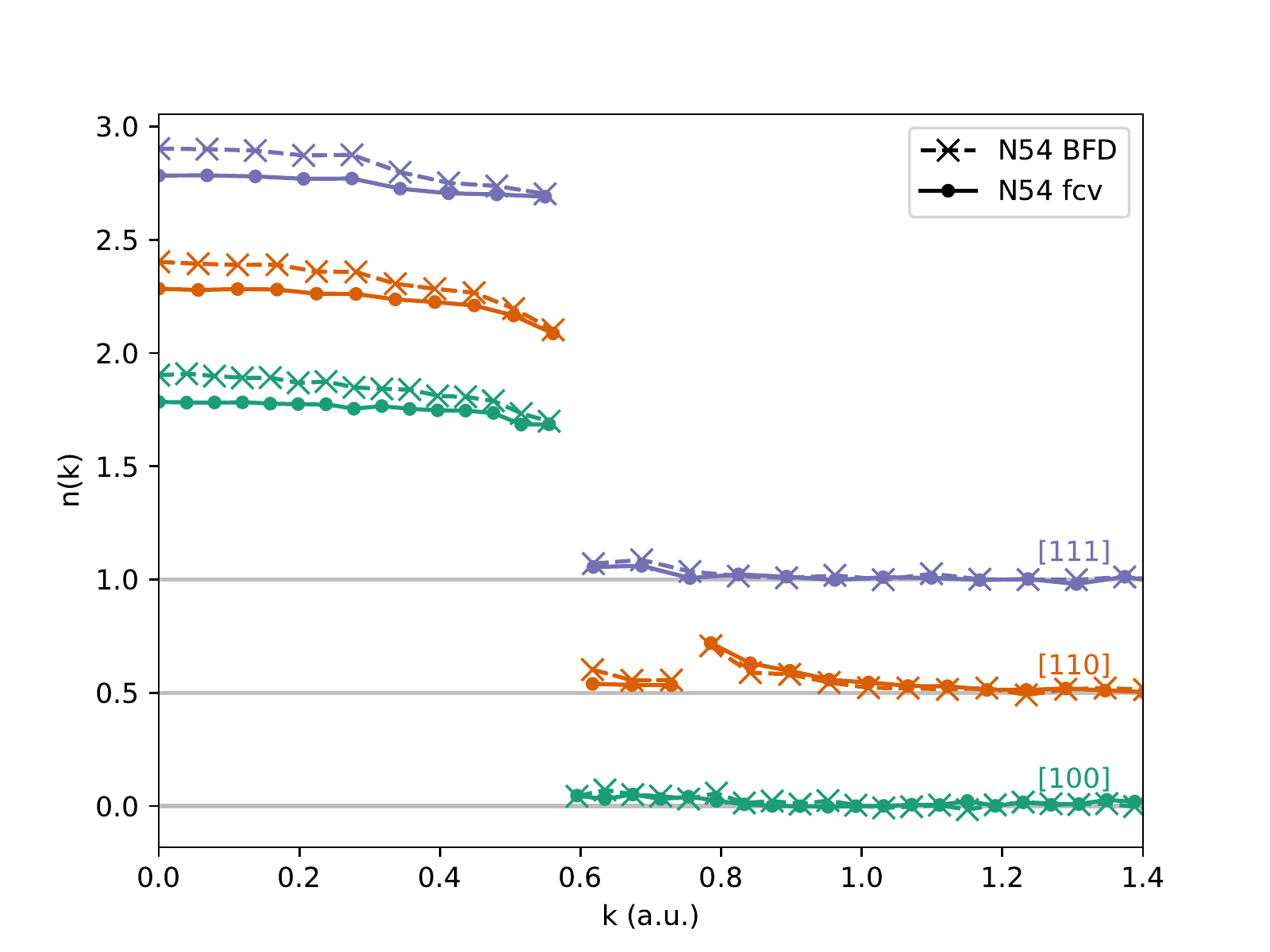}
(a) full-core valence \textit{vs} pseudopotential
\end{minipage}
\begin{minipage}{\columnwidth}
\includegraphics[width=\linewidth]{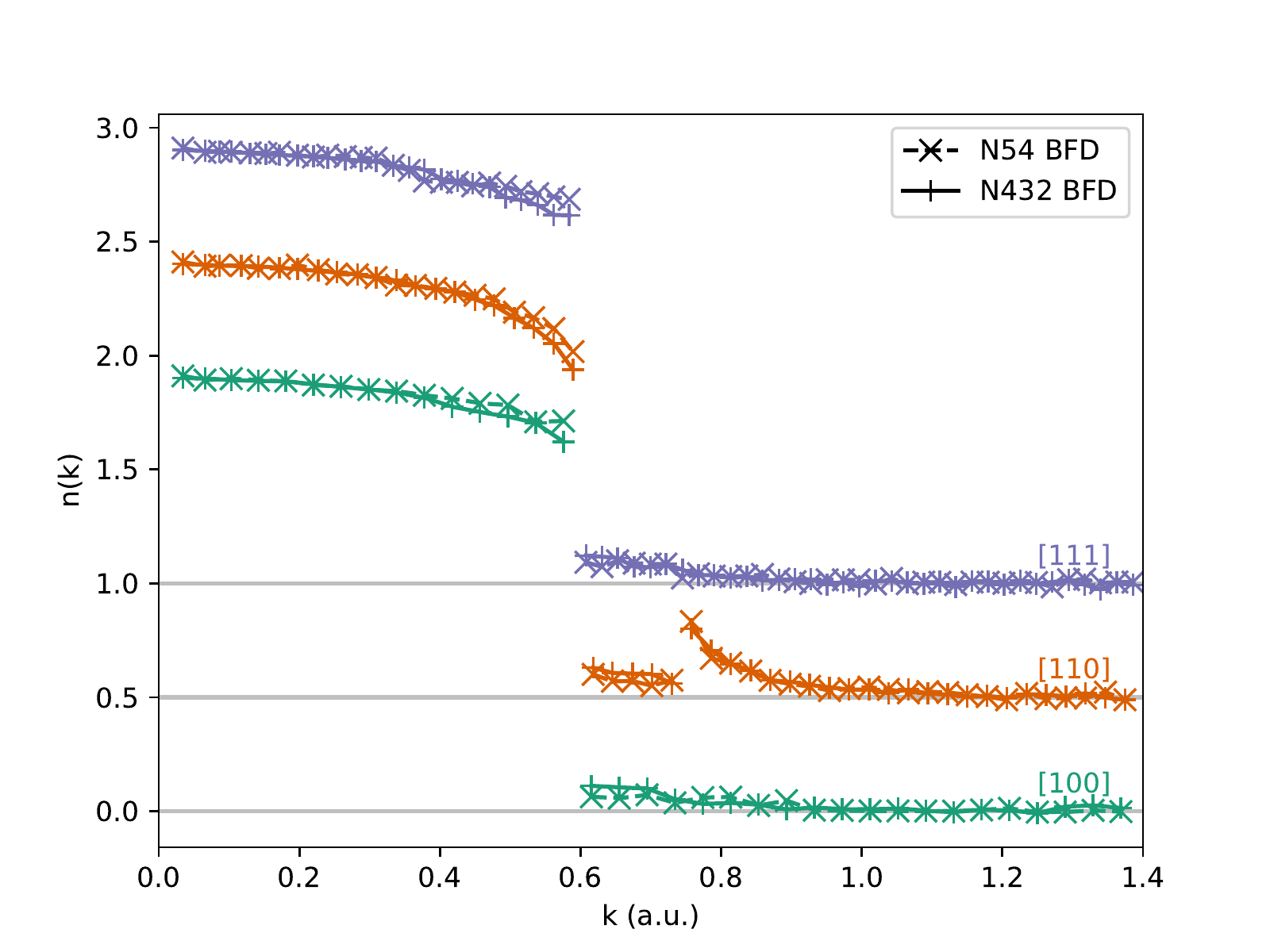}
(b) 432 atoms \textit{vs} 54 atoms
\end{minipage}
\caption{Momentum distribution of valence electrons in lithium BCC crystal. The top panel compares pseudopotential (crosses) to full-core (dots) result.
The bottom panel compares 54-atom (crosses) to 432-atom (pluses) pseudopotential results.\label{fig:nk}}
\end{figure}

In Fig.~\ref{fig:nk}, 1D slices of the QMC valence momentum distributions are shown. The momentum distribution is free-electron-like along the [100] and [111] directions. Along the [110] direction, however, there is a pronounced secondary Fermi surface. The valence profile from the full-core calculation is flatter inside the Fermi surface and has enhanced secondary features when compared to the pseudopotential calculation. 

To obtain the valence momentum distribution from the full-core QMC calculation, we remove the momentum distribution of the 1s core electrons. The 1s orbital of the neutral lithium atom is calculated using Hartree-Fock (HF) with a cc-pV5Z basis. The most pronounced effect of the pseudopotential is to increase the electronic momentum density inside the Fermi surface, raising $n(0)$ by more than 5\%. In contrast, the effect of increasing system size peaks at the Fermi momentum. The main effect of finite system size is to increase the magnitude of the discontinuity at the Fermi momentum. The effects of pseudopotential and finite system size can be better shown in the momentum distribution differences.

\begin{figure}
\begin{minipage}{\columnwidth}
\includegraphics[width=\linewidth]{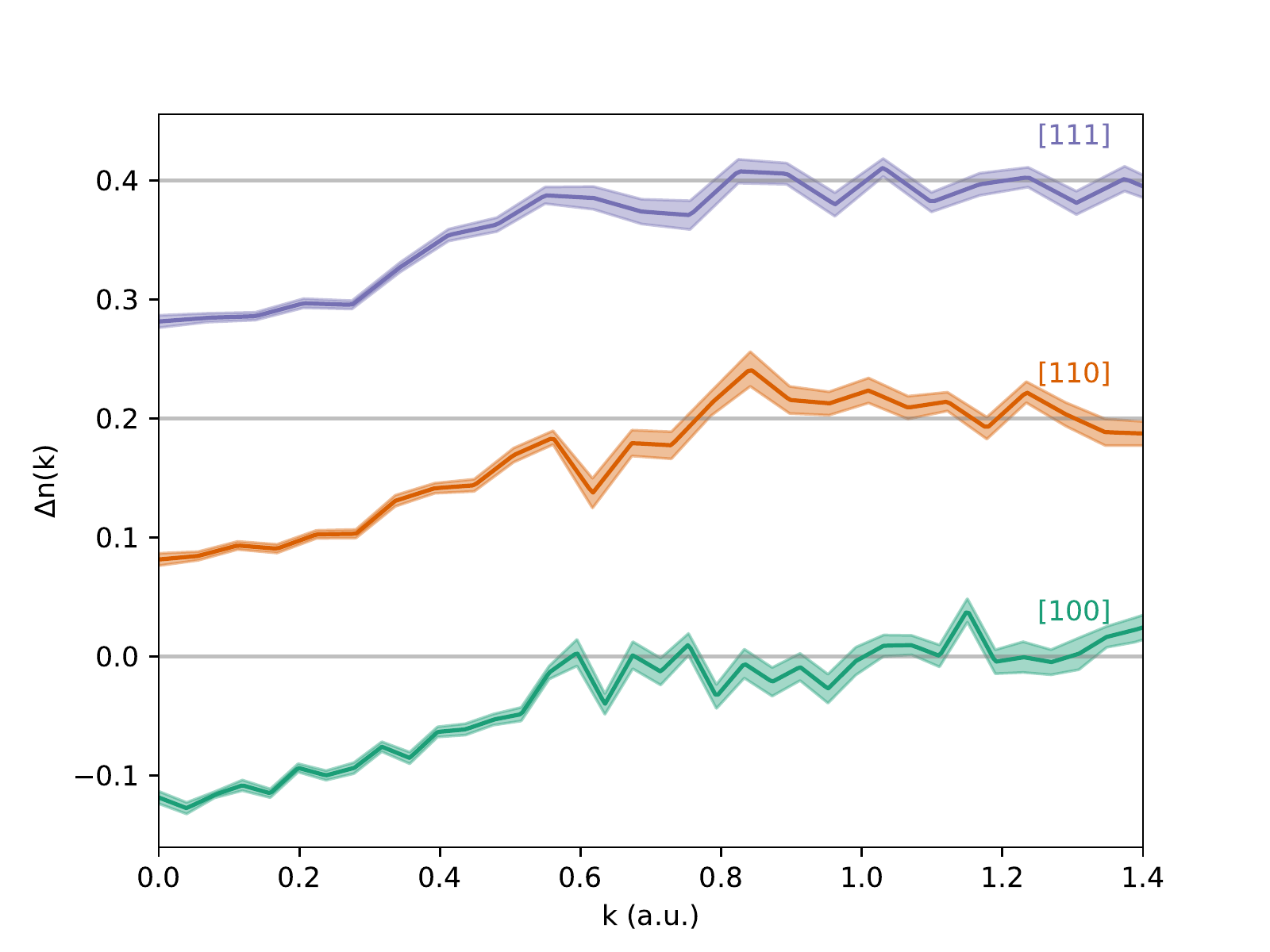}
(a) full-core valence - pseudopotential
\end{minipage}
\begin{minipage}{\columnwidth}
\includegraphics[width=\linewidth]{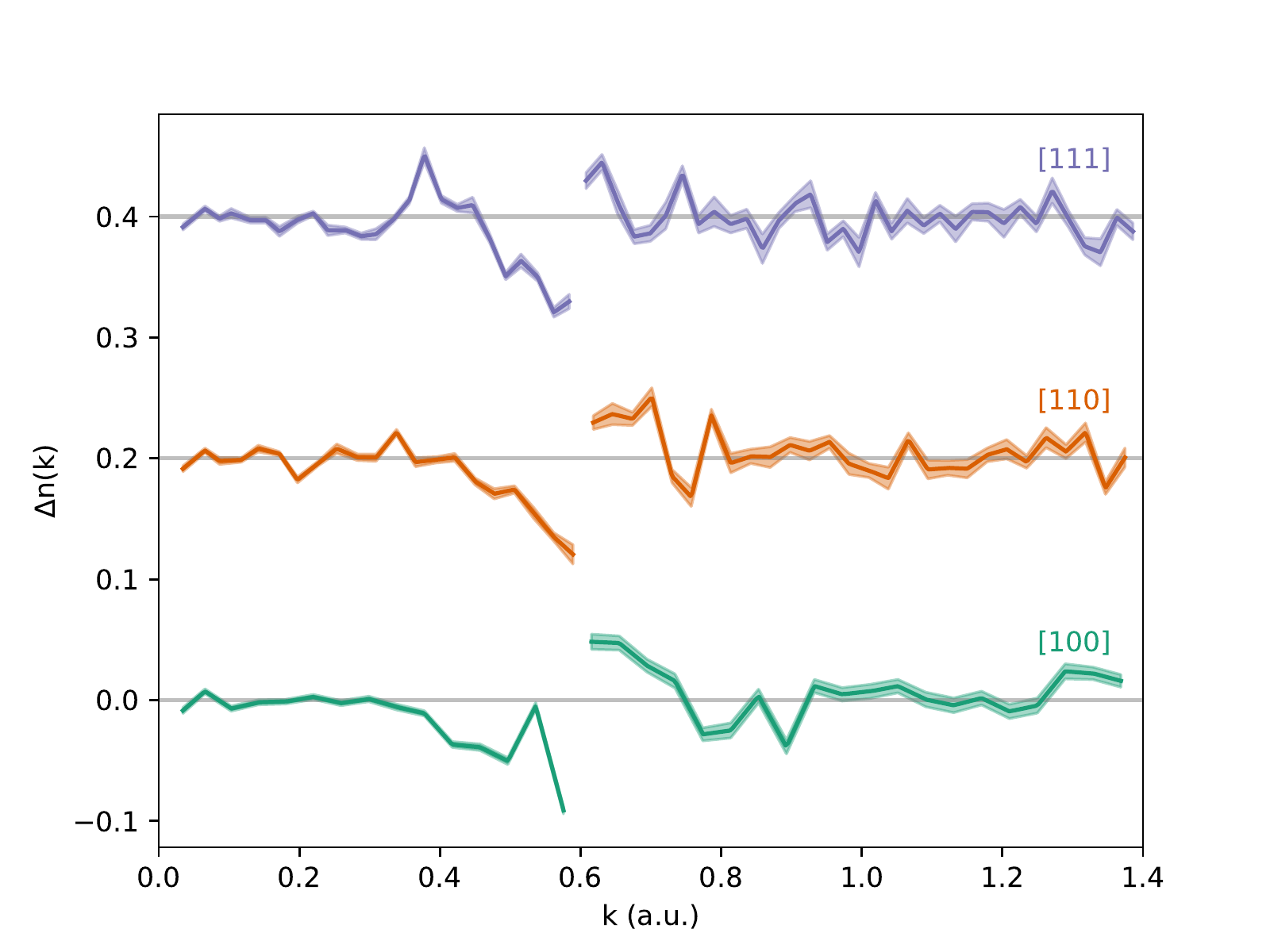}
(b) 432 atoms - 54 atoms
\end{minipage}
\caption{Momentum distribution differences. The top panel is the difference between full-core and pseudopotential results. The bottom panel is the difference between the 432-atom and 54-atom pseudopotential results. The shaded region show one standard deviation of statistical uncertainty. These results are used to inform pseudopotential and finite-size corrections. \label{fig:dnk}}
\end{figure}

In Fig.~\ref{fig:dnk}, we show two sets of momentum distribution differences in direct correspondence with Fig.~\ref{fig:nk}. The first is the difference between full-core and pseudopotential momentum distributions. This difference can be considered a pseudopotential correction (PPC). The PPC is largest inside the Fermi surface. It has a parabolic shape and is mostly negative along the [100] and [111] directions. However, it shows positive peaks near the secondary Fermi surface along the [110] direction.  The PPC is spherically-averaged and applied to the momentum distributions of the disordered structures.

Now consider how the finite size of our supercell affects the results: the finite-size correction (FSC). Figure~\ref{fig:dnk}(b) shows the difference between the 432-atom and 54-atom pseudopotential calculations. The difference peaks at the Fermi surface and goes to zero at high momenta.  The FSC results shown here are used to validate the approach outlined in ref.~\cite{Holzmann2009} and ref.~\cite{Holzmann2011}.

\begin{figure}
\includegraphics[width=\linewidth]{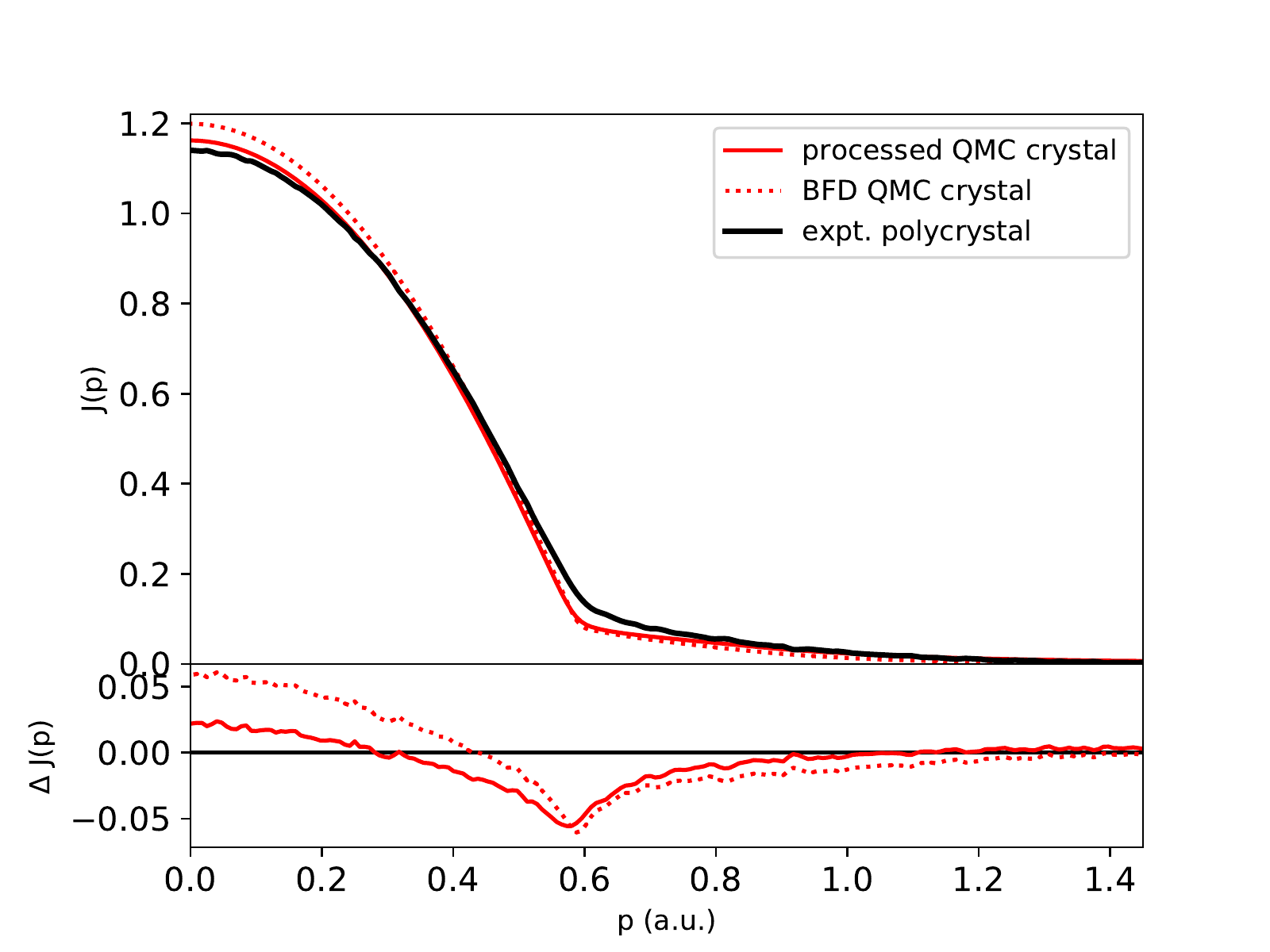}
\caption{Spherical average of the valence Compton profile of lithium BCC crystal at $r_s=3.25$. The red solid line is the best QMC result with all processing steps applied. The red dotted curve is our pseudopotential QMC result. The black curve is experiment on polycrystal lithium. \label{fig:crystal-vcp}}
\end{figure}

In Fig.~\ref{fig:crystal-vcp}, we show our best QMC Compton profile in the crystal as the red line. It is the spherically-averaged Compton profile from the 432-atom pseudopotential calculation with PPC and FSC applied. Further, we rescaled the QMC data to change density from $r_s=3.25$ to $r_s=3.265$ and convolved the QMC Compton profile with Eq.~(\ref{eq:elorentz}) to approximately account for experimental resolution and final-state effects. The full-core QMC profiles agrees well with the most recent experiment away from the Fermi surface.

The Compton profile reported by Filippi and Ceperley \cite{Filippi1999} is closer to our full-core than to our pseudopotential result. This is because they accounted for proper core-valence orthogonalization using full-core LDA. Pseudopotential QMC was used to estimate the correlation correction, rather than directly provide the Compton profile.

\begin{figure}
\includegraphics[width=\linewidth]{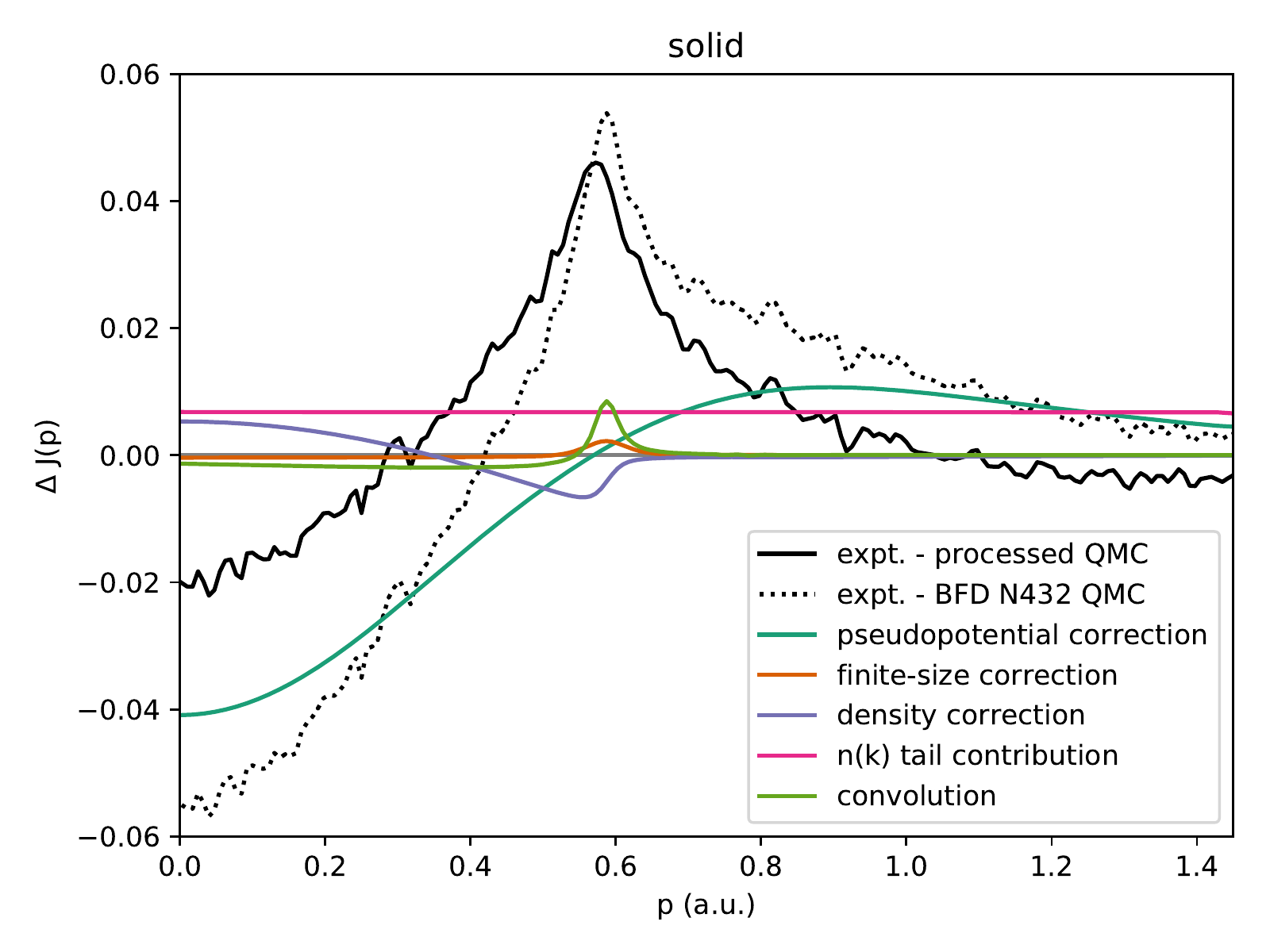}
\includegraphics[width=\linewidth]{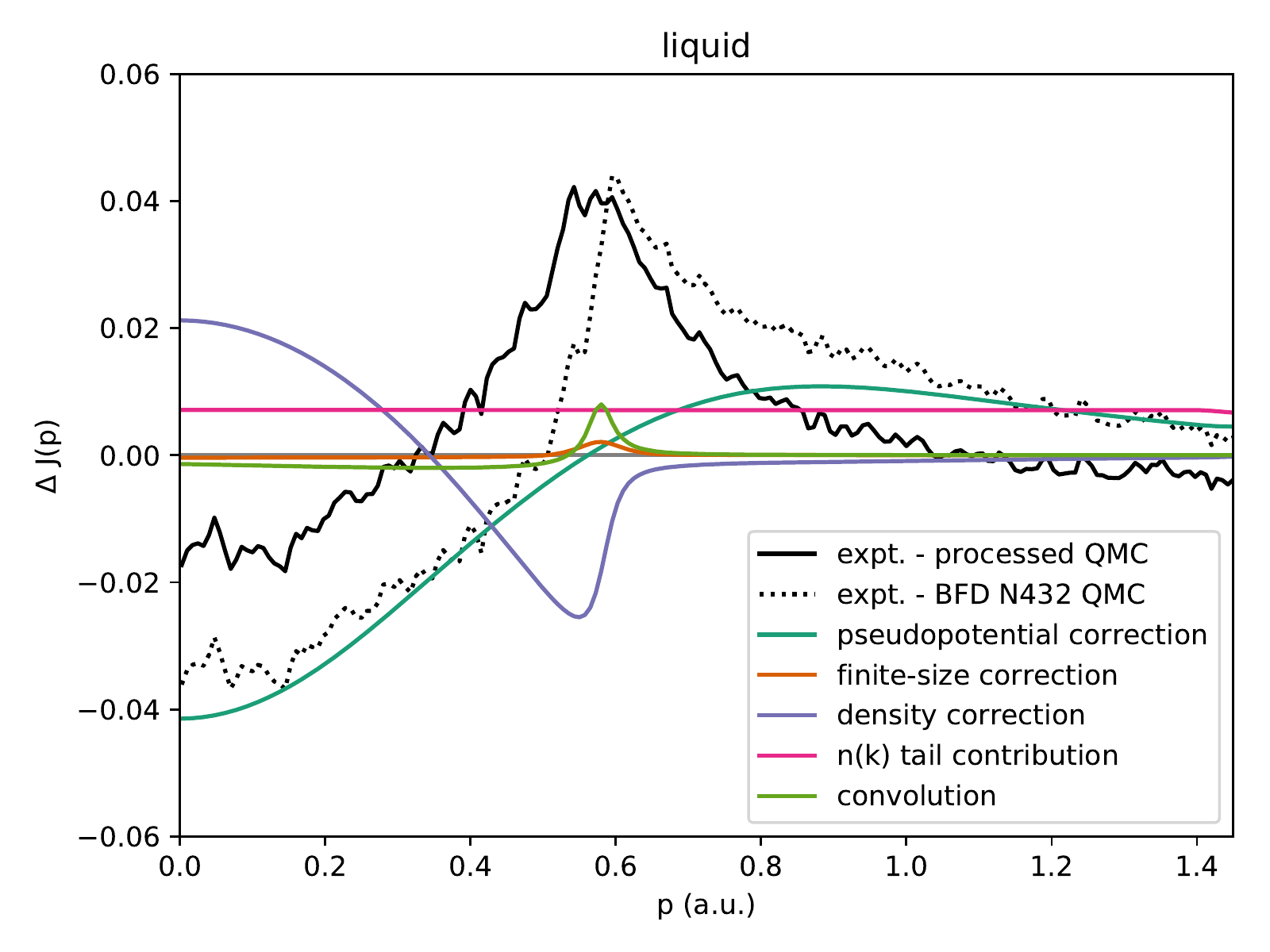}
\caption{Valence Compton profile corrections. The solid black curve is experiment relative to ``best'' theory. The dotted black curve is experiment relative to pseudopotential QMC result with no correction. Each colored curve shows the effect of neglecting a processing step from the theoretical Compton profile. When added to the processed result (solid black curve), the sum of all colored curves approximately recovers the unprocessed result (dotted black curve). \label{fig:sl-corrections}}
\end{figure}

Taking our best QMC Compton profiles (thin lines in Fig.~\ref{fig:sl-jp-djp}) as reference, we show the remaining difference between the QMC and the experiment Compton profiles as the black curves in Fig.~\ref{fig:sl-corrections}. We also show the effect of each processing step in the calculation of $J(p)$. Finite-size and convolution corrections both peak at the Fermi momentum and are small at the scale of the remaining discrepancy. The density correction is small in the solid but substantial in the liquid, because QMC calculations have been performed close to the solid density. In both cases, the density correction contracts the Fermi sphere and has little effect above the Fermi momentum. In contrast, the pseudopotential correction nearly vanishes at the Fermi momentum, smoothly transfers low-momentum components to high momenta, and remains non-zero well above the Fermi momentum. The $n(k)$ tail correction is needed to recover the normalization sum rule, because the QMC $n(k)$ is truncated at a finite momentum $k_c$. The exact shape of $n(k)$ tail may not be accurate above $k_c$, because the assumed functional form is simple (see supplemental materials). Fortunately, the effect of $n(k)$ tail within $k_c$ is simply to shift the entire Compton profile up by a constant as dictated by the normalization sum rule. The tail and pseudopotential corrections are the only ones that can change the high-momentum tail of the Compton profile.

\section{Discussion} \label{sec:discussion}

In the following, we discuss possible explanations for the remaining discrepancy in Fig.~\ref{fig:sl-jp-djp},
which is shown separately for the solid and liquid in Fig.~\ref{fig:sl-corrections}.

{ \bf Electron-Ion interaction} The crystal lattice introduces inhomogeneity to an otherwise homogeneous valence electron density. Umklapp processes send electronic momentum density to secondary Fermi surfaces, thereby enhancing the high-momentum components of the momentum distribution and reducing the momentum distribution inside the Fermi surface. Further, its discontinuity at the Fermi surface is reduced~\cite{Eisenberger1972}.
In the absence of other interactions, the ground-state electronic density will be exact if the electron-ion interaction is perfectly captured. DFT is designed to obtain the correct ground-state electronic density, so we expect it to treat electron-ion interaction well. However, pseudopotential is not designed to faithfully reproduce the charge inhomogeneity of the valence orbital in the core region. Therefore, pseudopotential introduces a bias in the valence momentum distribution.

\begin{figure}
\includegraphics[width=\linewidth]{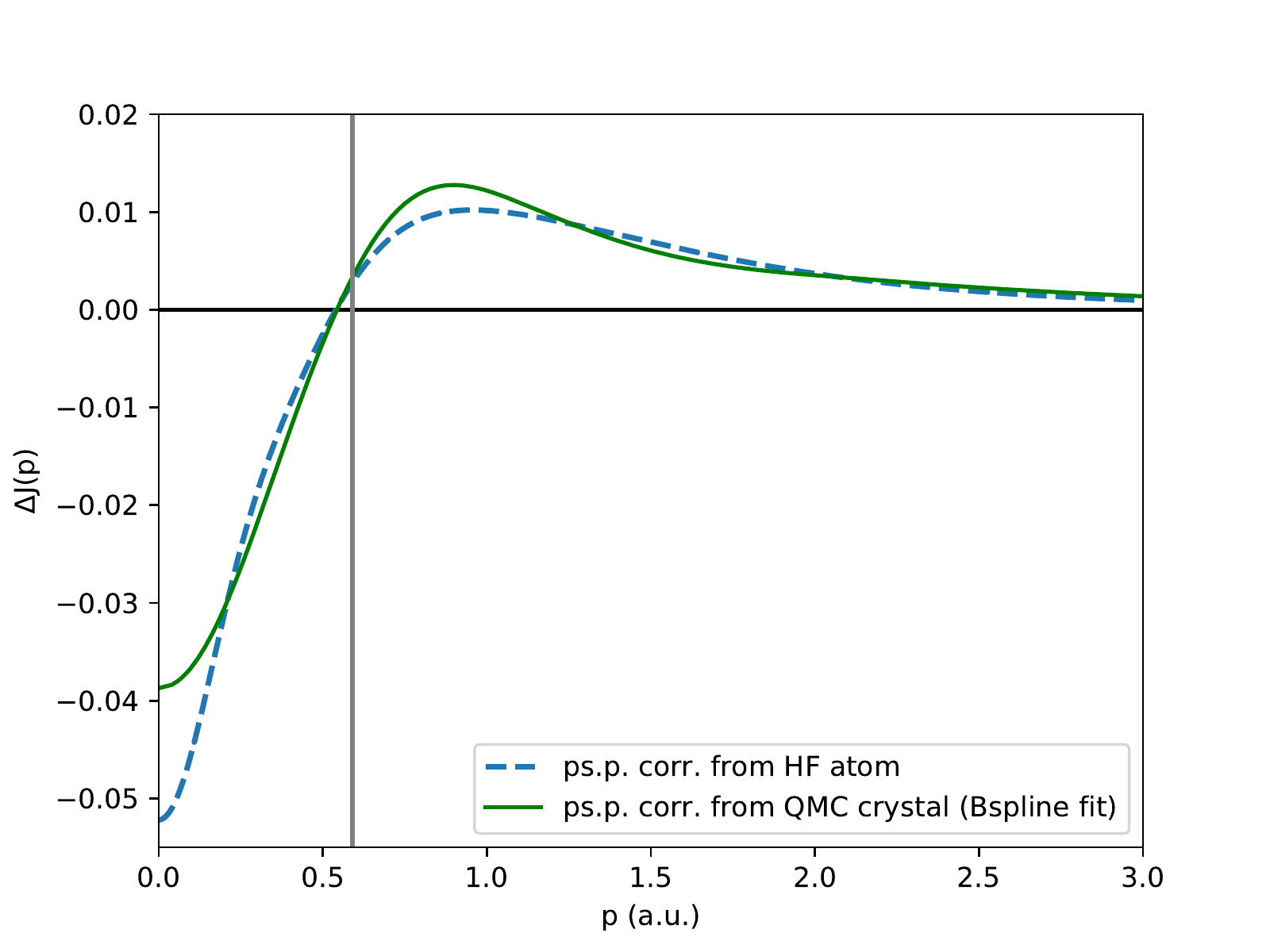}
\caption{Pseudopotential correction derived from QMC and HF. The green curve is the same QMC pseudopotential correction as shown in Fig.~\ref{fig:sl-corrections}. The dashed blue curve is the pseudopotential correction derived from the all-electron v.s. pseudized lithium atom using HF. The gray vertical line marks the Fermi momentum. \label{fig:hf-ppc}}
\end{figure}

The qualitative effect of the pseudopotential is clear from its construction. When designing a pseudopotential, one smooths the valence orbital inside the core region. This will decrease the electronic momentum density at high momenta, and increase it at low momenta. Indeed, one can reproduce the pseudopotential correction semi-quantitatively by considering the smoothing of the pseudized valence orbital in the lithium atom (Fig.~\ref{fig:hf-ppc}). We see that augmented planewave (APW) calculations~\cite{Baruah1999,Bross2004,Bross2005,Bross2012} tend to reproduce the experimental Compton profiles better at low momenta than pseudopotential calculations.

Our pseudopotential correction (PPC) is not perfect. It was derived in the perfect crystal, then applied to the disordered configurations. Ideally, one would directly perform all-electron QMC on the disordered configurations. However, this is computationally expensive. We do not consider all-electron calculation to be necessary in the solid phase, because the effect of disorder is small. The current PPC does over correct the liquid Compton profile at high momenta, because the corrections meant for the secondary Fermi surfaces are extraneous. Nevertheless, we think the pseudopotential bias is mostly captured, i.e. at the scale of Fig.~\ref{fig:hf-ppc}. The corrected Compton profile in Fig.~\ref{fig:crystal-vcp} is in better agreement with experiment than its pseudopotential counterpart, especially at $p=0$. We do not think the pseudopotential bias is responsible for the remaining discrepancy, because the PPC is concentrated around $p=0$. If it were underestimated, then the remaining correction would lower $J(0)$ much more than it would raise $J(p_F)$, worsening the agreement with experiment. 

{\bf Disorder}  Disorder mostly reduces the effect of the crystal lattice, because deviations from the perfect lattice weaken Umklapp processes. A confirmation was obtained when Sternemann et al. reproduced the temperature effect on the Compton profile of lithium by smearing out the pseudopotential with a Debye-Waller factor~\cite{Sternemann2001}.

Thermal disorder is also unlikely to be responsible for the remaining discrepancy because disorder-correction is small at the scale of the remaining correction. This can be seen by comparing the discrepancy in the perfect crystal (Fig.~\ref{fig:crystal-vcp}) to the discrepancy in the disordered solid (Fig.~\ref{fig:sl-jp-djp}). The two remaining discrepancies are similar in both shape and magnitude.

{\bf Electron-Electron Correlation}  The effect of electron-electron (ee) correlation on the momentum distribution is similar to electron-ion interaction in that it increases high-momentum components, decreases low-momentum components and reduces the discontinuity at the Fermi surface. The Slater-Jastrow wavefunction is a first-order modification of the free-electron Slater determinant by the Coulomb interaction~\cite{Holzmann2003} but it does not capture all correlation effects.
However, we expect the Slater-Jastrow wavefunction to be accurate for simple metals. 
Further, it can be systematically improved, for example by using backflow transformations \cite{PhysRevB.91.115106}. 
Calculations on the homogeneous electron gas indicate a small decrease of the discontinuity
at the Fermi 
surface \cite{Holzmann2011} reducing the discrepancy with experiment. Quantitative studies of
backflow effects on the lithium Compton profiles should be addressed in the future.

{\bf Fermi surface}  The Fermi surface of BCC lithium is anisotropic with pronounced secondary features. The DFT Fermi surface is used in the QMC simulation to determine which momentum states to occupy.
For solid lithium, the Fermi surface is nearly spherical. Our DFT Fermi surface of the BCC crystal has a maximum anisotropy of $\delta=5.0\%$, where
\begin{equation} \label{eq:ani-delta}
\delta \equiv \dfrac{k_F^{[110]} - k_F^{[100]}}{k_F^{HEG}}.
\end{equation}
This is in good agreement with the de Haas-van Alphen experiment performed by M. B. Hunt et al.~\cite{Hunt1989}, which reported a maximum anisotropy of $\delta=4.8\pm 0.3\%$. Our DFT result differs from previous calculations by A. H. MacDonald $\delta=3.3\%$~\cite{MacDonald1980} and H. Bross $\delta=5.9\%$~\cite{Bross2005}, likely due to differences in the density functional and pseudopotential.
While the DFT Fermi surface may not be accurate in the crystal,  a liquid is isotropic and will have a spherical Fermi surface.
Given that our solid - liquid Compton profile difference agrees well with experiment (Fig.~\ref{fig:s-l-djp}), we do not consider Fermi surface shape to be responsible for the remaining discrepancy.

{\bf Electron-phonon interaction} We capture disorder effects due to phonons by averaging over thermal atomic configurations. However, other phonon effects are absent from our QMC simulations because the lithium ions are clamped. Phonons scatter quasi-particles and decrease their life times. Thus, we expect the inclusion of electron-phonon interaction to decrease the magnitude of the discontinuity in the momentum distribution. Calculations of the coupled electron-phonon system within the Einstein or Debye model~\cite{PhysRev.131.993} show that the resulting broadening
at zero temperature is essentially given by the Debye frequency.
The Debye temperature of lithium ($<$400K) is much lower than the Fermi temperature of the electrons, so we expect the remaining electron-phonon coupling (not included
in our QMC calculations) to be limited very close to the Fermi surface in momentum space, rendering the effect invisible at the scale of Fig.~\ref{fig:s-l-djp}.

\begin{figure}
\includegraphics[width=\linewidth]{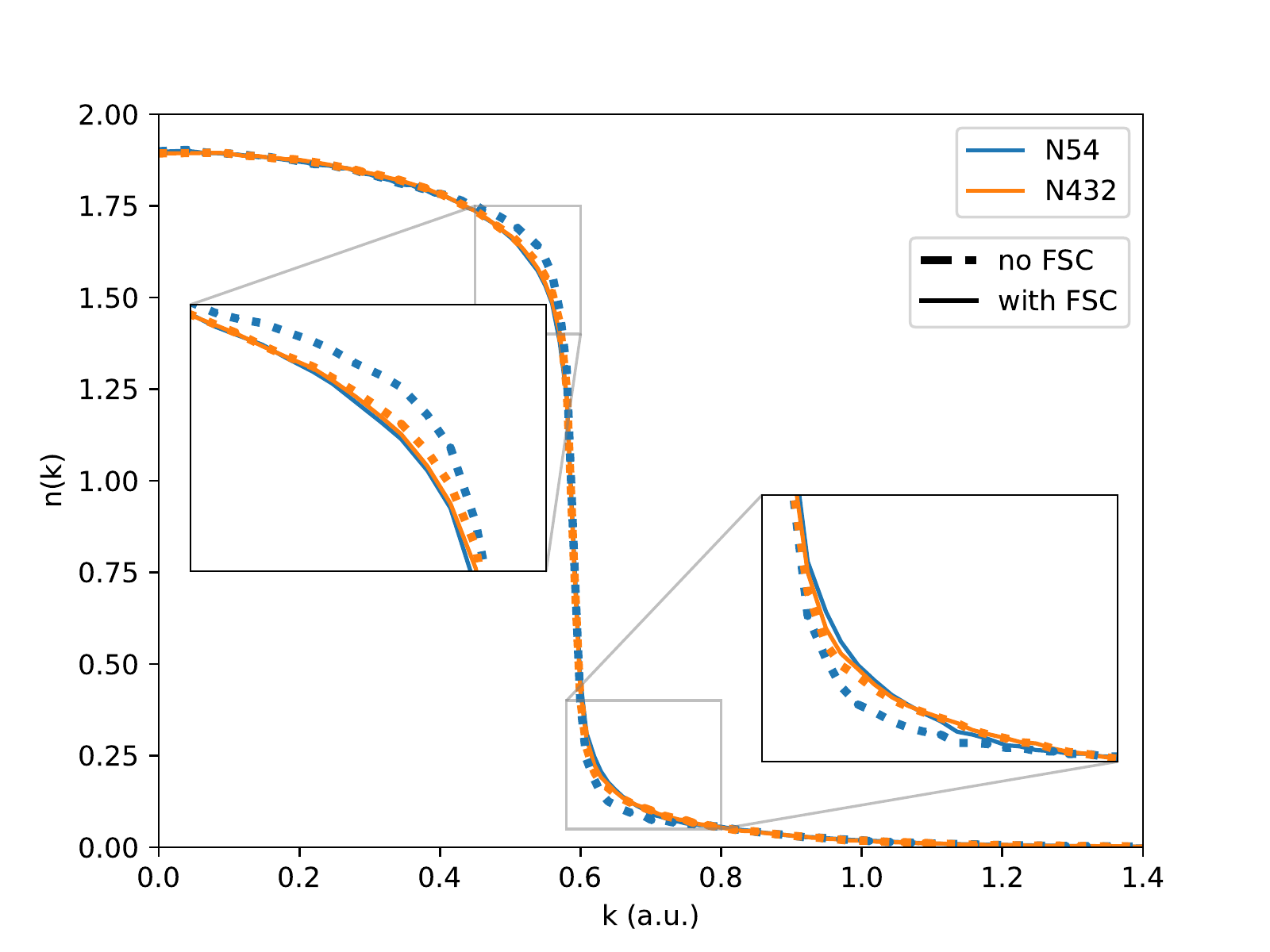}
\caption{Finite-size correction in the liquid phase. Dotted lines are pseudopotential QMC $n(k)$ with no correction. Color encodes the number of lithium atoms in the simulation cell. The solid lines correspond to the dotted lines in color and have been corrected using the leading-order expression Eq.~(\ref{eq:nk-fsc}). \label{fig:liquid-nk-fsc}}
\end{figure}

{\bf Finite size effects}  Finite-size effects (FSE) are more challenging to deal with in a many-body simulation than in an effective one-particle theory such as DFT which is formulated for an infinite lattice. In DFT, a calculation performed in a larger simulation cell simply makes the momentum-space grid denser. In contrast, finite system size increases the magnitude of the discontinuity at the Fermi surface in QMC. This effect was found to decrease slowly with system size in the homogeneous electron gas~\cite{Holzmann2009}. This FSE  was analyzed and understood in the homogeneous electron gas~\cite{Holzmann2009,Holzmann2011}. We adopted the same approach here and found good results. In particular, we corrected the FSE using the leading-order expression
\begin{equation} \label{eq:nk-fsc}
\delta n_{\bs{k}}^{(1)} = \int_{-\pi/L}^{\pi/L} \frac{d^3\bs{q}}{(2\pi)^3} \left[
u_q (1-S_q) - nu_q^2 S_q
\right] (n_{\bs{k}+\bs{q}}-n_{\bs{k}}),
\end{equation}
where $u_q$ and $S_q$ are the Jastrow pair function and the structure factor in reciprocal space, which are assumed to take RPA forms at small $q$ and $n$ is the valence electron density. The corrected $n(k)$ from the 54-atom and 432-atom simulations agree well with each other as shown in Fig.~\ref{fig:liquid-nk-fsc}. Therefore, we think finite-size error has been satisfactorily accounted for, and is not responsible for the remaining discrepancy.

{\bf Density change} The electronic density is a crucial parameter since it determines the Fermi surface. It can change due to thermal expansion and phase transition from solid to liquid.
We accounted for density change between our calculations and experiment by rescaling our computed momentum distributions to the experimental densities by scaling the value of $k$ to match the Fermi momentum ($k_F=(9\pi/4)^{1/3}/r_s$) and then correcting the overall normalization.
This brought the Compton profile into excellent agreement with experiment as shown in Fig.~\ref{fig:s-l-djp}.  Of course it would be possible to perform additional QMC simulations at the experimental density. 

{\bf Final state effects} Finally, the ``impulse approximation'' is known to be inaccurate for core electrons and cause asymmetry in the measured Compton profile~\cite{Eisenberger1970,Sternemann2000,Huotari2001}. To go beyond the ``impulse approximation'', one must consider interaction of the scattered electron with the rest of the system in the final state. Final-state effects are often attributed to three physical interactions. The first is the interaction between the excited quasi-particle with its surrounding medium (self-energy). The second is the interaction between the excited quasi-particle and the hole it lefts behind (vertex correction). The third is the interaction between the hole and a plasmon (plasmaron). C. Sternemann et al. showed that the self-energy combined with the vertex correction can satisfactorily explain the asymmetry of the Compton profile~\cite{Sternemann2000}. The effect of final-state interaction on the Compton profile can be approximated by convolving the spectral density function (SDF) of the excited electron with the ground-state Compton profile~\cite{Soininen2001}. This convolution smears out the derivative-discontinuity of the Compton profile at the Fermi momentum. Thus the convolution correction also peaks at the Fermi momentum.

We account for final-state effects by convolving the QMC Compton profiles with the broadening function Eq.~(\ref{eq:elorentz}), which is an accurate representation of the convolution of the experimental resolution function and the SDF obtained by Soininen et al.~\cite{Soininen2001}. However, the SDF in ref.~\cite{Soininen2001} did not include plasmaron or electron-hole effects. Further, we find near perfect agreement with experiment if the QMC profiles were broadened using a Lorentzian having FWHM $\Gamma=0.026$. In other words, if the neglected final-state effects were to introduce long tails into the SDF, then the QMC profiles would agree much better with experiment. Therefore, final-state effect is a plausible explanation for much of the remaining discrepancy.

\section{Conclusion and Outlook}

Leveraging new algorithms and hardware, we improved the QMC Compton profile of lithium and provided the first QMC results in the disordered solid and the liquid states. Our QMC Compton profiles agree very well with the most recent synchrotron experiment~\cite{Nozomu2019}. We resolved the discrepancy between pseudopotential QMC and experiment at zero and high momenta using an all-electron QMC calculation. We discussed potential explanations for the remaining discrepancy, which is concentrated at the Fermi surface. Future studies should consider final-state effects.

Current state-of-the-art QMC algorithms are ready to aid synchrotron experiments in understanding the measured Compton profiles. It would be interesting to revisit the challenging problem that is the 3D reconstructing of the momentum distribution from directional Compton profiles~\cite{Schulke1996,Tanaka2001}. Momentum resolution has been increased by new techniques in both theory and experiment. Further, all-electron QMC for lithium is feasible for perfect crystals in supercells containing thousands of electrons. The comparison between lithium and sodium will be particularly interesting, because they have the same crystal structure but very different electron-ion interactions~\cite{Eisenberger1972}. A detailed study of these systems can shed more light on the nature of electron-ion and perhaps the electron-phonon interactions in simple metals.

Finally, when sufficient accuracy has been achieved in both theory and experiment, one can study the difference between ground-state (QMC) and final-state (experimental) Compton profiles to extract information on the dynamic structure factor of the system.

\section{Acknowledgment}

YY and DMC were funded by DOE 0002911. This work made use of the Blue Waters sustained-petascale computing project and the Illinois Campus Cluster, supported by the National Science Foundation (awards OCI-0725070 and ACI-1238993), the state of Illinois, the University of Illinois at Urbana-Champaign and its National Center for Supercomputing Applications.
We would like to thank Ilkka Kyl\"anp\"a\"a and Jaron Krogel for valuable discussions.

\bibliographystyle{apsrev4-1}
\bibliography{main}

\end{document}


\title{Quantum Monte Carlo Compton profiles of solid and liquid lithium (Supplemental Materials)}
\author{Yubo Yang}
\affiliation{Department of Physics, University of Illinois, Urbana, Illinois 61801, USA}
\author{Nozomu Hiraoka}
\affiliation{National Synchrotron Radiation Research Center, Hsinchu 30076, Taiwan}
\author{Kazuhiro Matsuda}
\affiliation{Graduate School of Science, Kyoto University, Kyoto 606-8502, Japan}
\author{Markus Holzmann}
\affiliation{Univ. Grenoble Alpes, CNRS, LPMMC, 38000 Grenoble, France}
\affiliation{Institut Laue Langevin, BP 156, F-38042 Grenoble Cedex 9, France}
\author{David M. Ceperley}
\affiliation{Department of Physics, University of Illinois, Urbana, Illinois 61801, USA}
\maketitle

\subsection{Disordered Configurations}

\begin{figure}[h]
\includegraphics[scale=0.48]{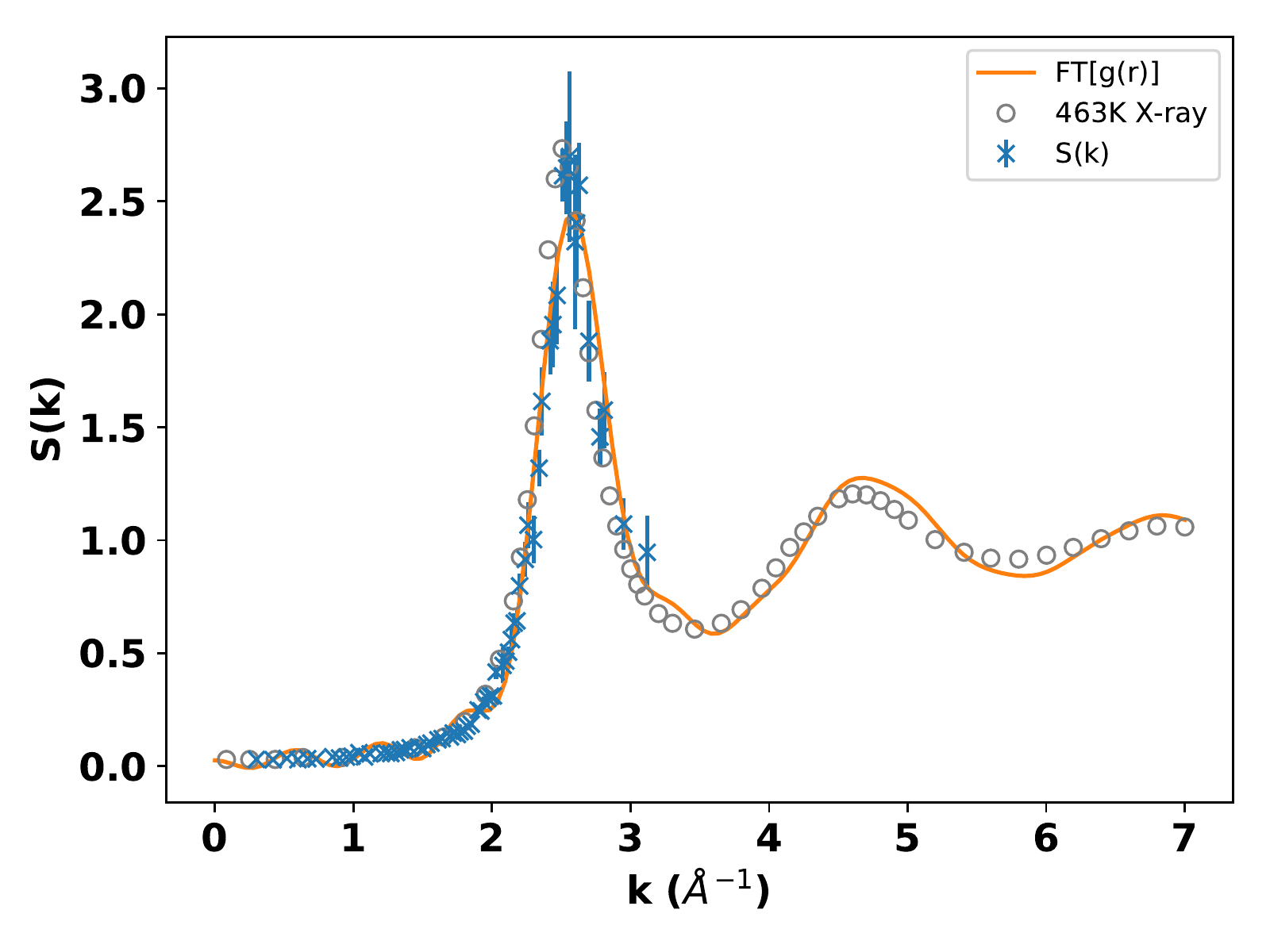}
\includegraphics[scale=0.48]{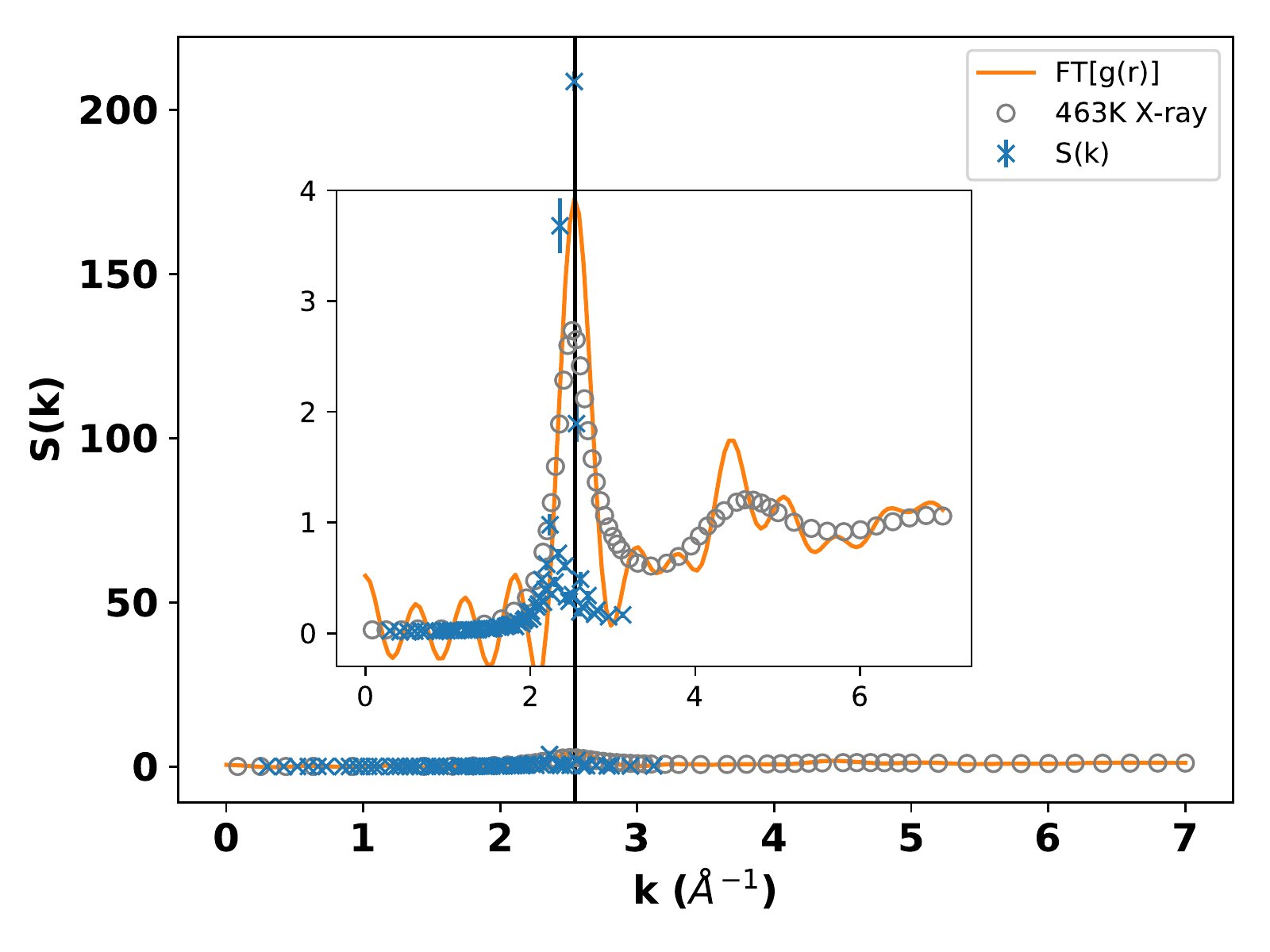}
\caption{Structure factor of disordered lithium configurations in liquid (left) and solid (right) phases. 32 lithium configurations were generated in each phase. Each configuration contained 432 lithium atoms in a cubic box with side length 20.96$\AA$. The liquid configurations were generated at $T=500K$, whereas the solid configurations were generated at $T=330K$. The blue crosses are spherically-averaged $S(k)$ calculated directly in reciprocal space. The solid line is the Fourier transform of the real-space pair correlation function $g(r)$. The gray circles are experimental values from X-ray scattering~\cite{Waseda1981,Mokshin2018}.}
\label{fig:lisk}
\end{figure}

Disordered configurations were generated from classical molecular dynamics (MD) simulations with the modified embedded atom potential (MEAM) as implemented in LAMMPS. 32 lithium configurations were generated in solid and liquid phases. The Li-Li structure factor was calculated from the MD runs and compared to X-ray data in Fig.~\ref{fig:lisk}.

\subsection{QMC Energies}

The orbitals in the Slater determinant were obtained using KS-LDA. We used a planewave cutoff of 256 Ry in the all-electron calculation. The resultant orbitals were modified to remove the approximate electron-ion cusp, which is exactly re-introduced in the Jastrow. All pseudopotential calculations used the BFD pseudopotential and a planewave cutoff of 16 Ry.

QMC calculations were carried out at valence density $r_s=3.25$, consistent with the previous study~\cite{Filippi1999}. 
QMC energies of the all-electron simulations are shown in the first three rows of Table~\ref{tab:qmc-etv}. Timestep error is $\sim$ 0.1 mha/e/a.u.. Mixed-estimator error of the kinetic energy is $\sim$ 4 mha/e. The mixed-estimator errors are larger than their pseudopotential counterparts. Nevertheless, our DMC total energy of -2.5152 ha/e is 2.3 mha/e lower than the -2.5129 ha/e obtained in a previous QMC study using localized basis and the PBE functional~\cite{Rasch2015}. The previous study was performed at a valence density of $r_s=3.24$, which is close to our $r_s=3.25$.
Energies from the pseudopotential simulations are shown in the remaining rows of Table~\ref{tab:qmc-etv}. The difference between pseudopotential DMC and VMC total energies are consistently around 1 mha/e. Further, the timestep error is $\sim$ 0.1 mha/e/a.u. and the mixed-estimator error of the kinetic energy is $<$ 1 mha/e. These small differences verify the high quality of our trial wavefunction for the valence electrons.

\begin{table}[h]
\caption{QMC energies and variance. All energies are reported in ha/e. Variance is in ha$^2$/e. Timestep is in ha$^{-1}$. Monte Carlo acceptance rate (acc) is in percent. Classical temperature is shown in Kelvin. $\langle\rangle$ indicates average over thermal ensemble and grand-canonical twist grid.}
\clearpage{}\begin{tabular}{rrrllllll}
\toprule
 $N_e/N_{Li}$ &  Classical T &  $\langle N_e\rangle$ & method & timestep &    acc & $\langle E\rangle/\langle N_e\rangle$ & $\sigma_E^2/\langle N_e\rangle$ & $\langle T\rangle/\langle N_e\rangle$ \\
\midrule
            3 &            0 &                161.93 &    VMC &      0.8 &  43.51 &                           -2.51138(2) &                       0.0204(2) &                              2.506(2) \\
            3 &            0 &                161.93 &    DMC &     0.01 &  98.79 &                           -2.51515(1) &                      0.01887(2) &                             2.5106(4) \\
            3 &            0 &                161.93 &    DMC &    0.005 &  99.53 &                           -2.51518(1) &                      0.01889(2) &                             2.5107(3) \\
            1 &            0 &                 54.11 &    VMC &        2 &  81.88 &                           -0.25682(3) &                      0.00393(4) &                            0.15041(7) \\
            1 &            0 &                 54.11 &    DMC &      0.2 &   98.7 &                           -0.25818(1) &                      0.00431(2) &                            0.14940(3) \\
            1 &            0 &                 54.11 &    DMC &      0.1 &  99.47 &                          -0.258074(8) &                      0.00420(1) &                            0.14951(3) \\
            1 &            0 &                431.41 &    VMC &     3.25 &  71.64 &                          -0.254623(3) &                     0.004175(9) &                           0.150825(8) \\
            1 &            0 &                431.41 &    DMC &      0.2 &   98.7 &                          -0.255837(4) &                     0.004632(8) &                            0.15022(2) \\
            1 &            0 &                431.41 &    DMC &      0.1 &  99.47 &                          -0.255737(4) &                     0.004520(7) &                            0.15029(2) \\
            1 &          330 &                431.85 &    VMC &        3 &  73.91 &                        -0.2520638(10) &                     0.004274(3) &                           0.152855(3) \\
            1 &          330 &                431.85 &    DMC &      0.3 &  97.85 &                         -0.2534244(9) &                     0.004872(2) &                           0.152209(3) \\
            1 &          330 &                431.85 &    DMC &     0.15 &  99.11 &                         -0.2532869(9) &                     0.004803(2) &                           0.152276(3) \\
            1 &          500 &                431.90 &    VMC &        3 &  73.54 &                          -0.249701(1) &                     0.004383(3) &                           0.154635(3) \\
            1 &          500 &                431.90 &    DMC &      0.3 &  97.81 &                         -0.2511173(9) &                     0.005000(2) &                           0.154009(3) \\
            1 &          500 &                431.90 &    DMC &     0.15 &  99.09 &                         -0.2509847(9) &                     0.004937(2) &                           0.154079(3) \\
\bottomrule
\end{tabular}
\clearpage{}
\label{tab:qmc-etv}
\end{table}

\subsection{QMC Electronic Structure Factor}

\begin{figure}[h]
\includegraphics[width=0.48\linewidth]{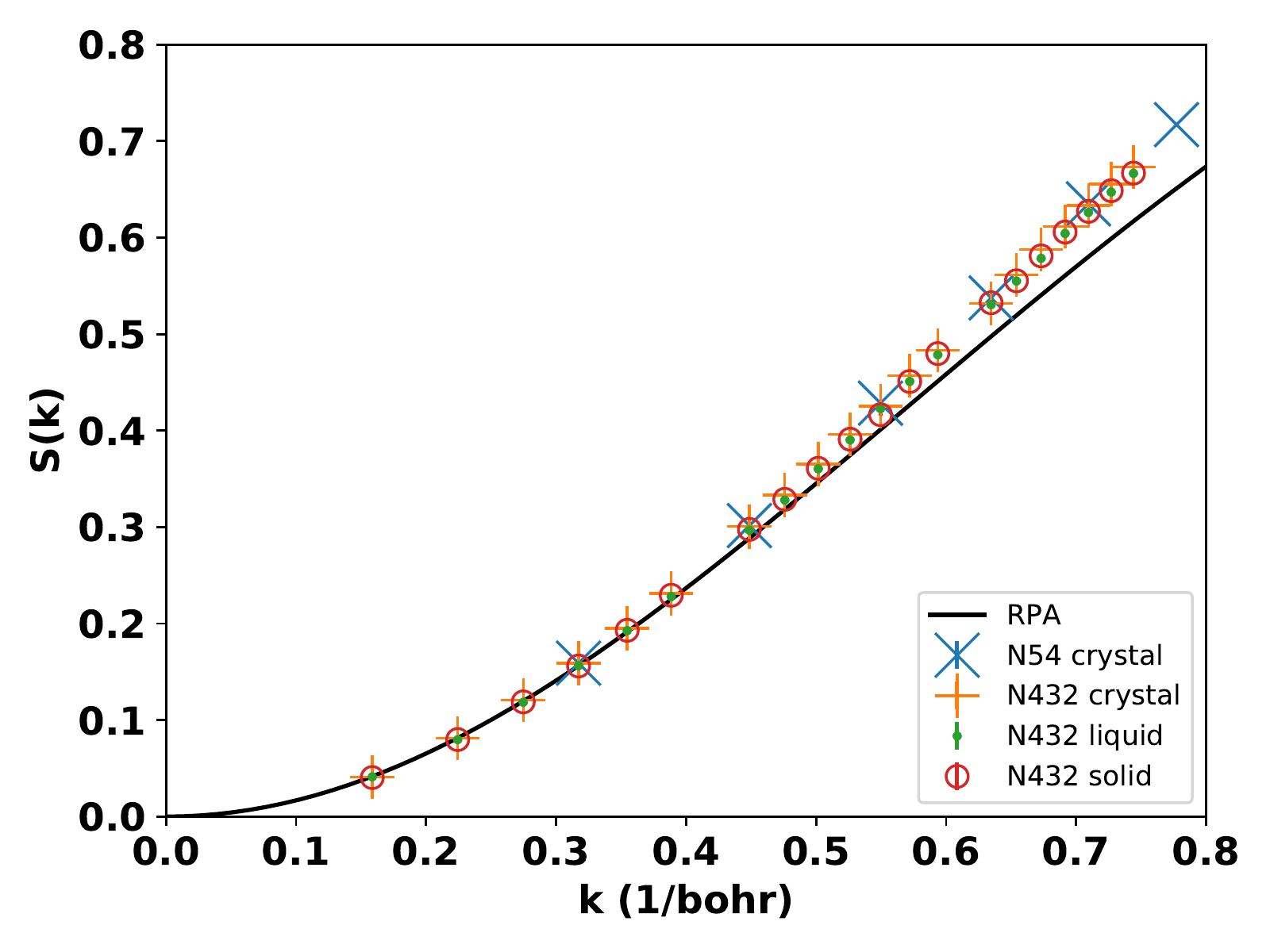}
\includegraphics[width=0.48\linewidth]{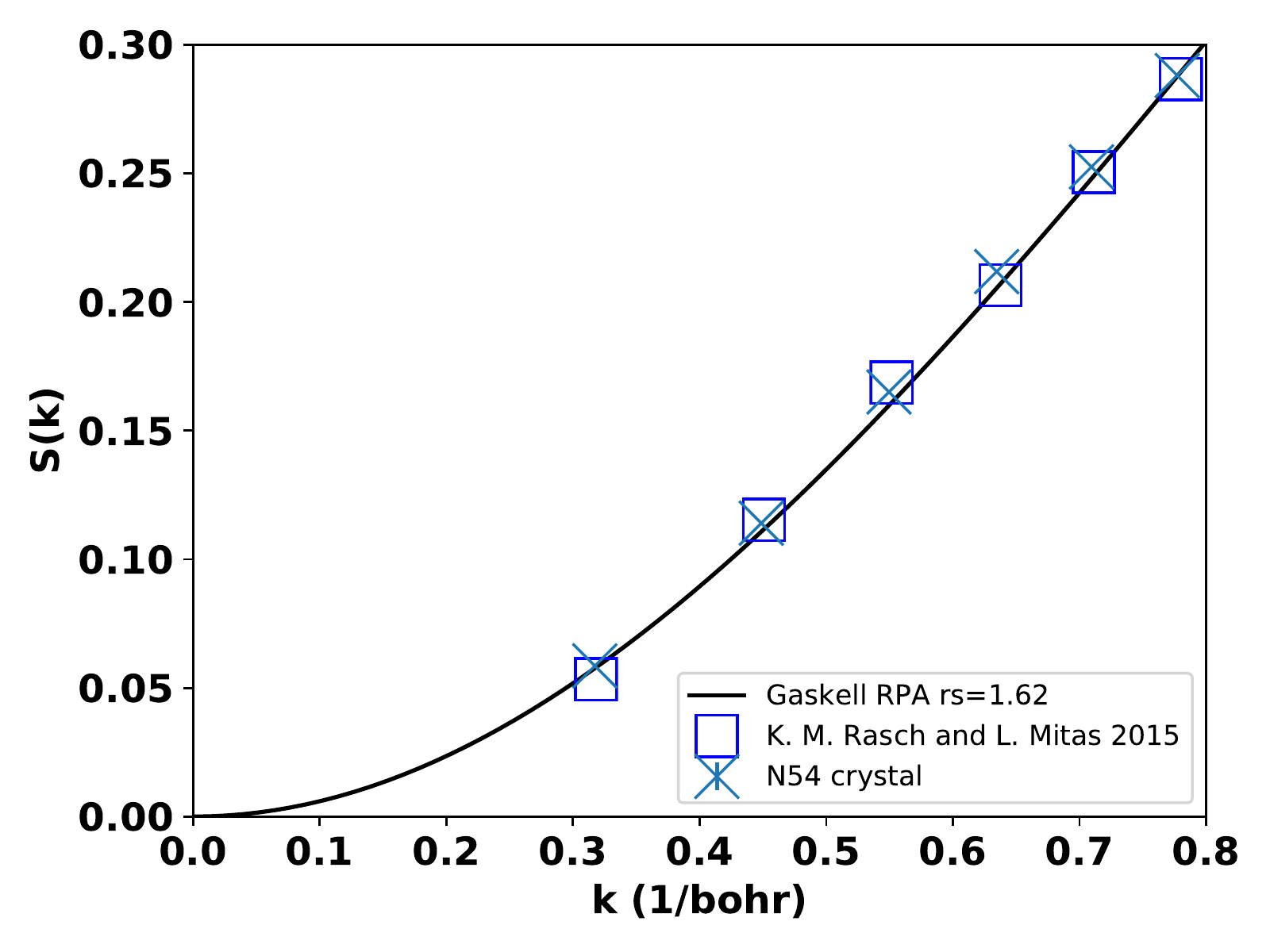}
\caption{Electronic static structure factor of pseudopotential (left) and all-electron (right) QMC simulations in 54-atom and 432-atom simulation cells. The black line in the left plot is RPA $S(k)$ at valence density $r_s=3.25$. It fits lithium valence $S(k)$ remarkably well for $k<0.4$ bohr$^{-1}$. In the right plot, the black line is RPA $S(k)$ at density $r_s=3.25/\sqrt{3}$. \label{fig:qmc-dsk}}
\end{figure}

The fluctuating electronic structure factor
\begin{align}
\delta S(k) \equiv \left\langle
(\rho_{\bs{k}}-\bar{\rho}_{\bs{k}})^* (\rho_{\bs{k}}-\bar{\rho}_{\bs{k}})
\right\rangle,
\end{align}
where $\rho_{\bs{k}} = \sum\limits_j^{N_e} e^{i\bs{r}_j\cdot\bs{k}}$ is the collective coordinate of the electrons. The $\braket{}$ denotes expectation value, and $\bar{\rho}_{\bs{k}}\equiv\braket{\rho}_{\bs{k}}$. The QMC fluctuating structure factors are shown in Fig.~\ref{fig:qmc-dsk}. All values are linearly extrapolated to remove the mixed-estimator bias. The pseudopotential $\delta S(k)$ is insensitive to disorder. The $\delta S(k)$ from 432-atom simulations of perfect crystal, disordered solid, and liquid structures are indistinguishable from one another. Further, finite system size has only a small effect on the electronic structure factor, because the $\delta S(k)$ of the 54-atom simulation is very close to the 432-atom one. All pseudopotential $\delta S(k)$ can be accurately described by the RPA $S(k)$ at the same valence density when $k<0.4$ a.u.. Our all-electron structure factor agrees well with that from the previous QMC study~\cite{Rasch2015}.

\subsection{QMC Momentum Distribution}

The momentum distribution is obtained on a cubic regular grid with spacing $dk=0.040$ a.u. in reciprocal space. To achieve this grid spacing, uniform twist-average grids of size $8^3$ and $4^3$ are used in the 54-atom and 432-atom simulations, respectively. The twist-average grid is $\Gamma$-centered in the perfect crystal calculations and shifted by $dk/2$ in all directions in disordered calculations. In the perfect crystal, cubic symmetries reduce the number of unique twists from 64 to 4 and 512 to 20 on a shifted grid, and 64 to 10 and 512 to 35 on a $\Gamma$-centered grid. The reciprocal-space grid is truncated at a spherical cutoff of $1.49$ a.u. in the solid and liquid simulations.

\begin{figure}[h]
\includegraphics[width=0.8\linewidth]{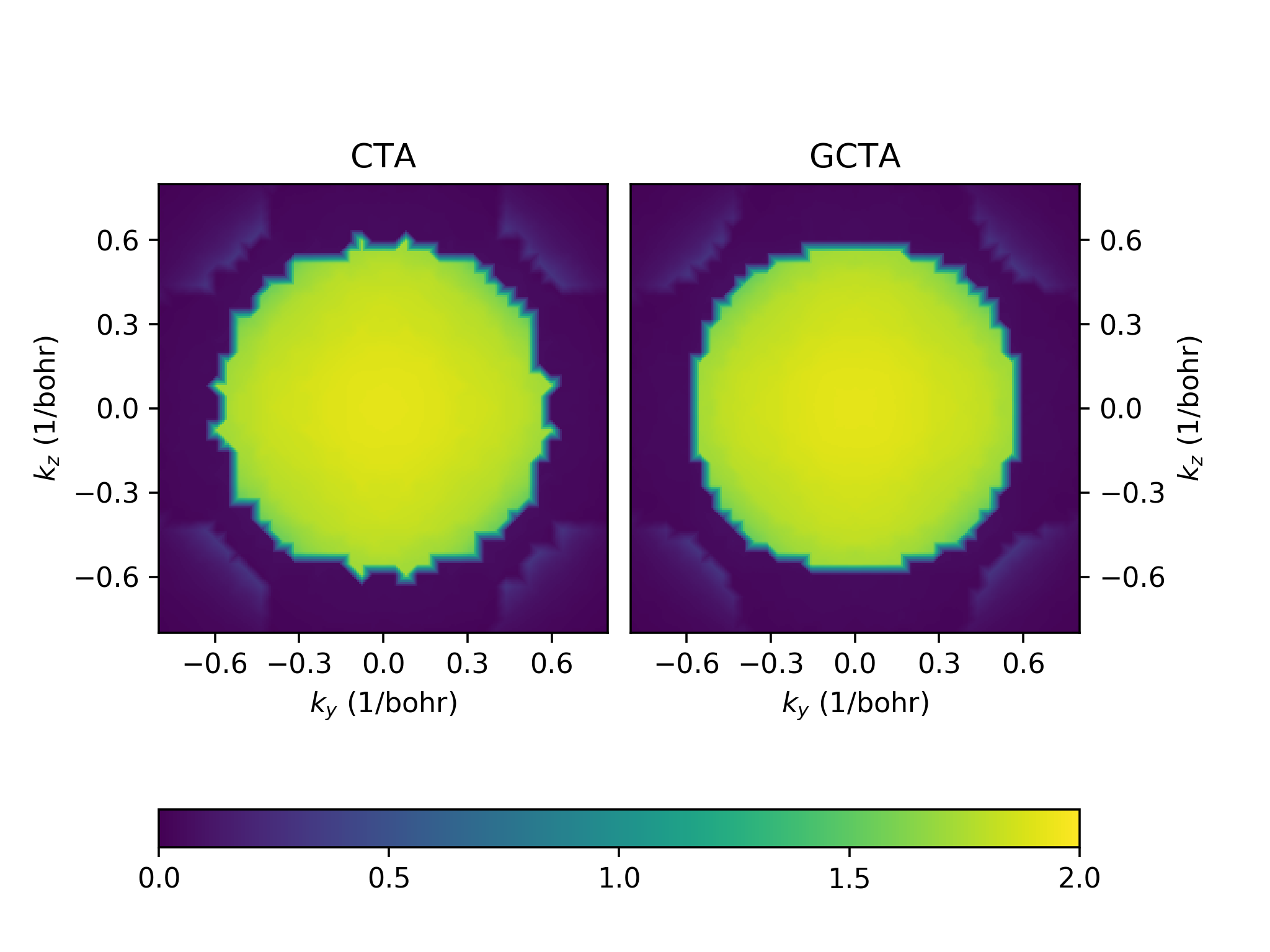}
\caption{[100] slice of n($\bs{k}$) at $k_x=0$ from canonical twist average grid (CTA) and grand-canonical twist average grid (GCTA). Occupation outside the Fermi surface can be seen as small extrusions along $k_y$ or $k_z$ in the canonical case. Both the main and secondary Fermi surfaces are visibly more circular in the grand-canonical case.}
\end{figure}

A QMC Compton profile is computed in four steps. First, the 3D DMC $n(\bs{k})$ is linearly extrapolated to reduce the mixed-estimator bias using VMC data. Second, the linearly-extrapolated $n(\bs{k})$ is spherically averaged to obtain a 1D distribution $n(k)$. Third, the 1D $n(k)$ is extended to large k using a model function eqn.~(\ref{eq:nktail}), which is inspired by the momentum distribution of hydrogen-like atoms
\begin{align} \label{eq:nktail}
n_{tail}(k, A, Z) = 2A\left(\dfrac{2Z}{(k^2+Z^2)^2}\right)^2,
\end{align}
where the $A$ and $Z$ are parameters are chosen to satisfy the sum rules as shown in Table~\ref{tab:crystal-ntsum}. Fourth and finally, the 1D $n(k)$ is integrated to obtain the spherically-averaged Compton profile $J(p)$ via eqn.~(\ref{eq:jp1d}) and split into two parts at a cutoff momentum $k_c$
\begin{align} \label{eq:jp1d}
J(p) = \left( \dfrac{(2\pi)^3}{\Omega/N_{Li}} \right)^{-1} \left[
\int\limits_p^{k_c} k n(k) dk + \int\limits_{k_c}^{\infty} k n_{tail}(k, A^*, Z^*) dk
\right],
\end{align}
where $\Omega$ is the volume of the supercell, $N_{Li}$ is the number of lithium atoms in the supercell, and $A^*$, $Z^*$ are chosen to satisfy momentum distribution sum rules. The sum-rules of the momentum distribution and the Compton profile in our convention are
\begin{align}
\left( \dfrac{(2\pi)^3}{\Omega/N_{Li}} \right)^{-1} \int d\bs{k} n(\bs{k}) =& \braket{N_e}/N_{Li}\equiv \bar{N}, \label{eq:nsum} \\
\left( \dfrac{(2\pi)^3}{\Omega/N_{Li}} \right)^{-1} \int d\bs{k} n(\bs{k})~\frac{\hbar^2k^2}{2m_e} =& \braket{T}/N_{Li} \equiv \bar{T}, \label{eq:tsum} \\
\int_{-\infty}^{\infty} dp J(p) =& \braket{N_e}/N_{Li}, \label{eq:jpsum}
\end{align}
where $\braket{T}$ is the expectation value of the total kinetic energy.
The fitted values in the tail function eqn.~(\ref{eq:nktail}) are shown in Tables~\ref{tab:crystal-ntsum}, along with sum-rule compliance.

\begin{table}[h]
\caption{Fits to $n(k)$ tails and sum rule compliance. $\bar{N}$ and $\bar{T}$ are the normalization and kinetic energy sum rules as defined in eqn.~(\ref{eq:nsum}) an (\ref{eq:tsum}). $\bar{N}_0$ and $\bar{T}_0$ are expected sum-rule values calculated from Table~\ref{tab:qmc-etv}. All sum-rule integrals are split into two parts at $k=k_c$ in the same way as eq.~(\ref{eq:jp1d}). $\Delta n(k_c)\equiv n_{tail}(k_c)-n(k_c)$ is the difference between the fit analytical tail and QMC data at the split point. The second row is ``full-core valence'' obtained by subtracting the HF core contribution from the QMC ae calculation. The HF core kinetic energy of the Li atom is 7.2239067 ha.}
\clearpage{}\begin{tabular}{rrrlrrrrrrrr}
\toprule
 $N_e/N_{Li}$ &  Classical T &  $N_{Li}$ &    $\bar{T}_0$ &  $\bar{T}$ &  $\bar{N}_0$ &  $\bar{N}$ &       A &       Z &  $k_c$ &  $\Delta n(k_c)$ &  $n(k_c)$ \\
\midrule
            3 &            0 &        54 &       7.539(2) &     7.5388 &       2.9986 &     2.9986 &  7.1015 &  2.7625 &   1.50 &          -0.0064 &    0.0519 \\
            3 &            0 &        54 &       0.315(2) &     0.3149 &       0.9986 &     0.9986 &  0.1831 &  2.9175 &   1.50 &          -0.0013 &    0.0023 \\
            1 &            0 &        54 &     0.14925(7) &     0.1488 &       1.0021 &     1.0030 &  0.0213 &  0.4958 &   1.45 &           0.0008 &    0.0006 \\
            1 &            0 &       432 &     0.14958(4) &     0.1494 &       0.9986 &     0.9989 &  0.0220 &  0.4959 &   1.45 &           0.0005 &    0.0009 \\
            1 &          330 &       432 &    0.151647(7) &     0.1519 &       0.9996 &     0.9992 &  0.0236 &  0.4961 &   1.45 &           0.0003 &    0.0013 \\
            1 &          500 &       432 &    0.153489(8) &     0.1539 &       0.9998 &     0.9990 &  0.0258 &  0.4964 &   1.45 &           0.0001 &    0.0015 \\
\bottomrule
\end{tabular}
\clearpage{}
\label{tab:crystal-ntsum}
\end{table}

\begin{figure}[h]
\begin{minipage}{0.49\linewidth}
\includegraphics[width=\columnwidth]{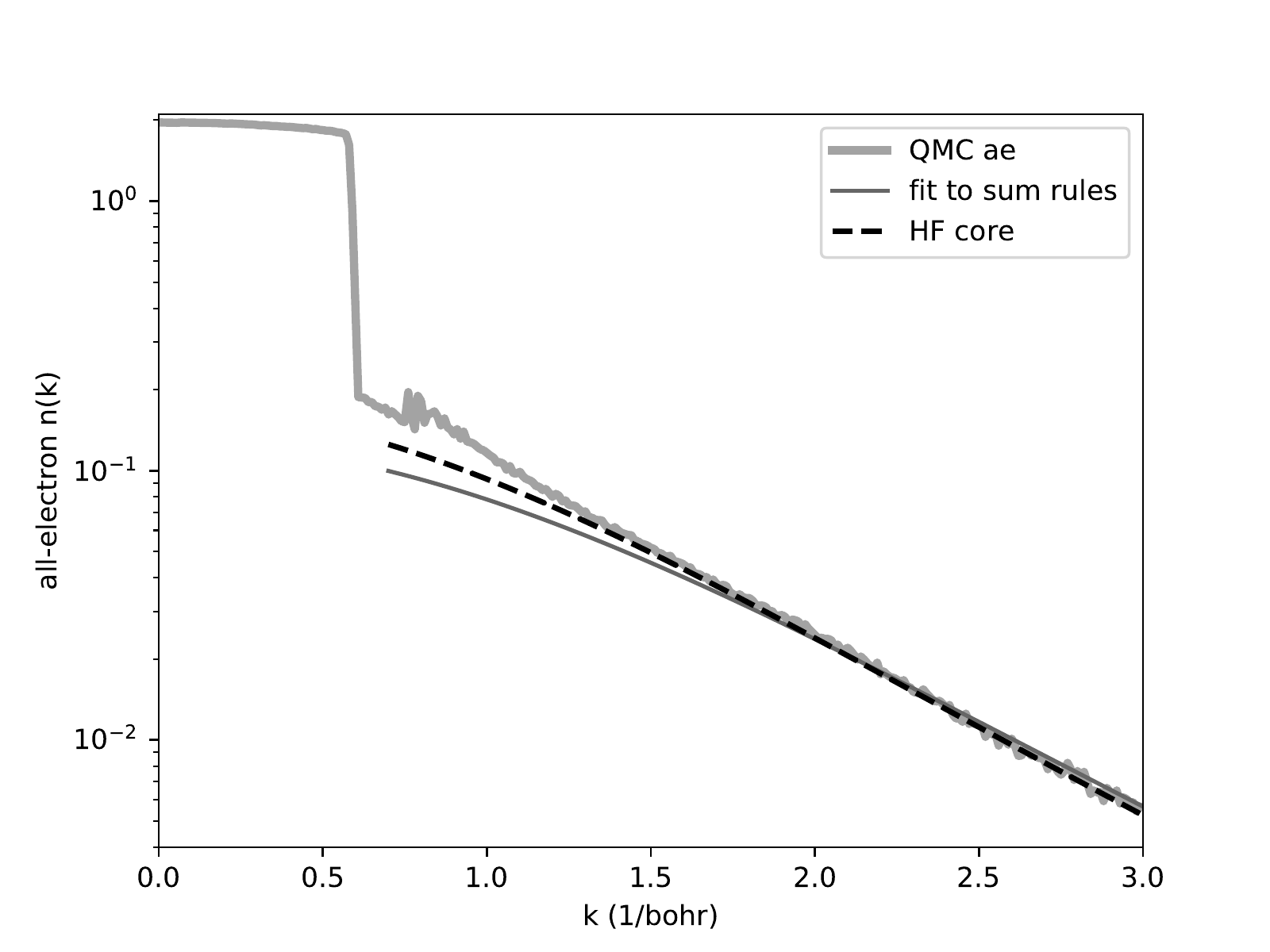}
(a) all-electron $n(k)$
\end{minipage}
\begin{minipage}{0.49\linewidth}
\includegraphics[width=\columnwidth]{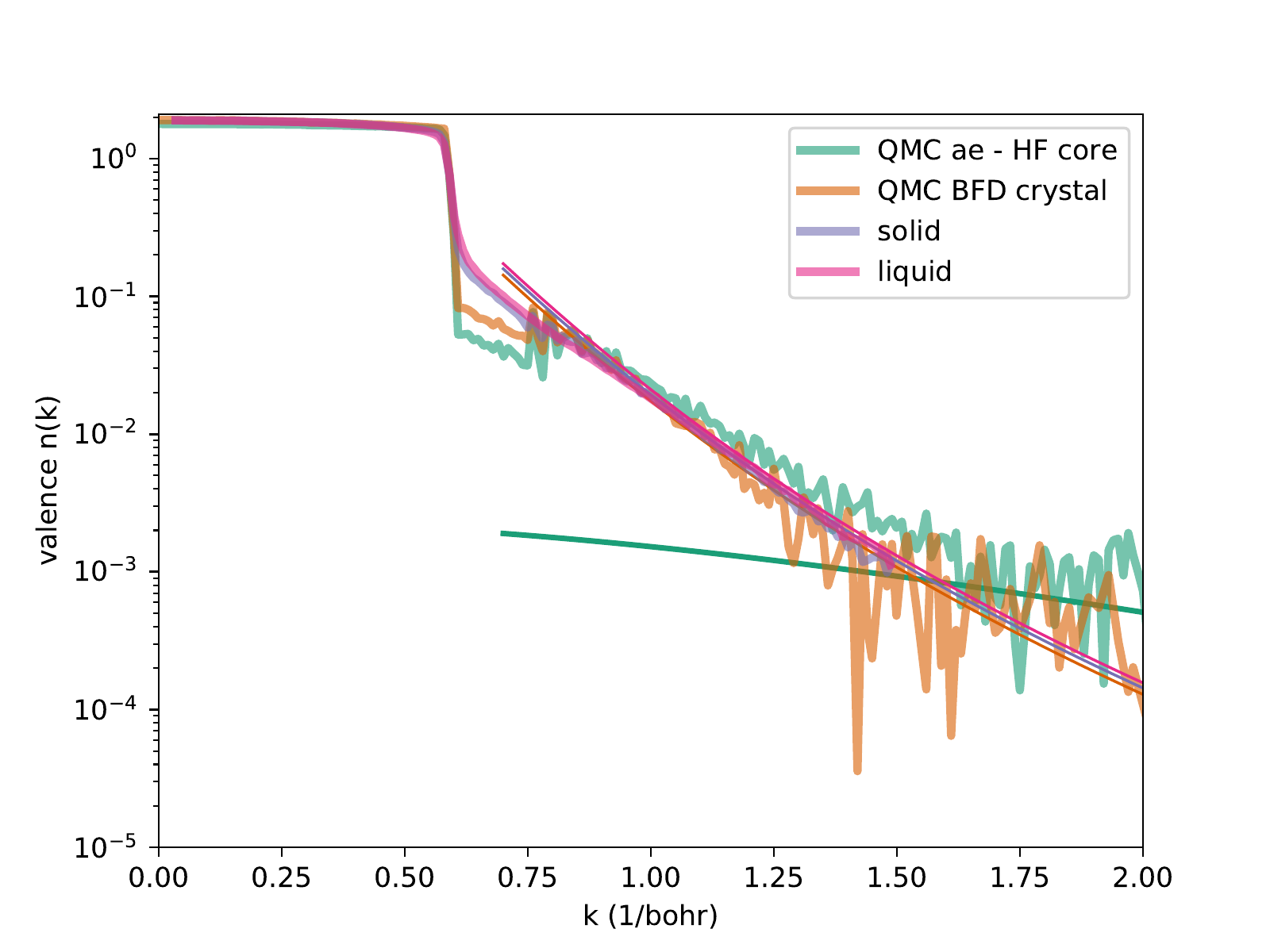}
(b) valence $n(k)$
\end{minipage}
\caption{Models of the high-momentum tail of $n(k)$ based on sum-rule compliance. Left plot shows all-electron momentum distribution from QMC (thick gray line) and Li atomic core contribution from HF (dashed black line). The thin gray line is the $n(k)$ tail model for QMC. Right plot shows valence momentum distributions from QMC (thick colored lines). Each thin line is the tail model for the QMC $n(k)$ of corresponding color. The statistical error of QMC $n(k)$ is on the order of $10^{-4}$ for $k>1.5$, so data close to $10^{-4}$ at large $k$ are not reliable.}
\label{fig:qmc-tail-model}
\end{figure}

Tail models in Table~\ref{tab:crystal-ntsum} are shown along with QMC $n(k)$ data in Fig.~\ref{fig:qmc-tail-model}. The QMC all-electron $n(k)$ for the crystal appears to decay slightly slower than HF core $n(k)$ at large $k$. This causes the full-core valence momentum distribution to have a much longer tail than the pseudopotential ones.

\bibliographystyle{apsrev4-1}
\bibliography{supp-main}